\documentclass[11pt]{article}
\pdfoutput=1
\usepackage{dcolumn}
\usepackage{bm}

\usepackage{graphicx}
\usepackage{amssymb,amsmath}
\usepackage{multirow}
\usepackage{cite,color,url}
\usepackage[colorlinks=true
,urlcolor=blue
,anchorcolor=blue
,citecolor=blue
,filecolor=blue
,linkcolor=blue
,menucolor=blue
,linktocpage=true
,pdfproducer=medialab
,pdfa=true
]{hyperref}

\usepackage{epsfig,psfrag,rotating,soul}
\usepackage{rotfloat}

\evensidemargin \oddsidemargin
\allowdisplaybreaks

\psfrag{lk}{$l_k$}
\psfrag{lm}{$l_m$}
\psfrag{blk}{$\bar {l}_k$}
\psfrag{blm}{$\bar {l}_m$}

\def\thefootnote{\fnsymbol{footnote}}
\def\e{\boldsymbol{e}}
\def\m{\boldsymbol{\mu}}
\def\t{\boldsymbol{\tau}}
\def\tme{\theta_{\mu e}}
\def\tte{\theta_{\tau e}}
\def\ttm{\theta_{\tau \mu}}

\def\cte{c_{\tau e}}
\def\ctm{c_{\tau \mu}}\def\t{\boldsymbol{\tau}} 

\begin{document}
\thispagestyle{empty}

\begin{flushright}
IFT-UAM/CSIC-14-018\\
FTUAM-14-016\\
\end{flushright}

\vspace{0.5cm}

\begin{center}

\begin{Large}
\textbf{\textsc{Imprints of massive inverse seesaw model neutrinos in lepton flavor violating Higgs boson decays}}
\end{Large}

\vspace{1cm}

{\sc
E. Arganda$^{1}$%
\footnote{\tt \href{mailto:ernesto.arganda@unizar.es}{ernesto.arganda@unizar.es}}%
, M.J. Herrero$^{2}$%
\footnote{\tt \href{mailto:maria.herrero@uam.es}{maria.herrero@uam.es}}%
, X. Marcano$^{2}$%
\footnote{\tt \href{mailto:xabier.marcano@uam.es}{xabier.marcano@uam.es}}%
, C. Weiland$^{2}$%
\footnote{\tt \href{mailto:cedric.weiland@uam.es}{cedric.weiland@uam.es}}%
}

\vspace*{.7cm}

{\sl
$^1$Departamento de F\'{\i}sica Te\'orica, Facultad de Ciencias,\\
Universidad de Zaragoza, E-50009 Zaragoza, Spain

\vspace*{0.1cm}

$^2$Departamento de F\'{\i}sica Te\'orica and Instituto de F\'{\i}sica Te\'orica, IFT-UAM/CSIC,\\
Universidad Aut\'onoma de Madrid, Cantoblanco, 28049 Madrid, Spain

}

\end{center}

\vspace*{0.1cm}

\begin{abstract}
\noindent
In this paper we consider a Higgs boson with mass and other properties compatible with those of the  recently discovered Higgs particle at the LHC,
and explore the possibility of new Higgs leptonic decays, beyond the standard model, with the singular feature of being lepton flavor violating (LFV).
We study these LFV Higgs decays, $H \to l_k\bar l_m$, within the context of the inverse seesaw model (ISS) and consider the most generic case where three additional
pairs of massive right-handed singlet neutrinos are added to the standard model particle content. We require in addition that the input parameters of this ISS model
are compatible with the present neutrino data and other constraints, like perturbativity of the neutrino Yukawa couplings. We present a full one-loop computation of
the BR($H \to l_k\bar l_m$) rates for the three possible channels, $l_k\bar l_m=\mu \bar \tau,\, e \bar \tau,\, e \bar \mu$, and analyze in full detail the predictions
as functions of the various relevant ISS parameters.  We study in 
parallel the correlated one-loop predictions for the radiative decays, $l_m \to l_k \gamma$, 
within this same ISS context, and require full compatibility of our predictions with the present experimental bounds for the three radiative decays,
$\mu \to e \gamma$,  $\tau \to \mu \gamma$, and $\tau \to e \gamma$. After exploring the ISS parameter space we conclude on the maximum allowed LFV Higgs decay rates within the ISS.
\end{abstract}

\def\thefootnote{\arabic{footnote}}
\setcounter{page}{0}
\setcounter{footnote}{0}

\newpage

\section{Introduction}
\label{intro}
At present,  there seems to be a broad consensus in the high energy physics community that the recently discovered scalar particle
at the CERN-LHC~\cite{Aad:2012tfa,Chatrchyan:2012ufa} behaves as the Higgs particle of the standard model of particle physics (SM).
The most recent measurements of this scalar particle mass by the ATLAS and CMS collaborations set $m_h^{\rm ATLAS}=125.5\pm 0.6 $ GeV~\cite{Aad:2013wqa}
and $m_h^{\rm CMS}=125.7\pm 0.4 $ GeV~\cite{Chatrchyan:2013lba}, respectively. These experiments also show that the most probable $J^P$ quantum numbers for this discovered
Higgs boson are $0^+$, and conclude that the measured Higgs particle couplings to the other SM particles are in agreement so far, although yet with moderate precision, with
the values predicted in the SM. Also the scalar Higgs-like particle width $\Gamma_h$ has been found to be $\Gamma_h <  17.4$ MeV which is about 4.2 times the SM value~\cite{Width}. 
 
On the other hand, there is also a major consensus that the SM must be modified in order to include the neutrino masses and oscillations in agreement with present data,
which are nowadays quite impressive and urge of an explanation from a theoretical framework beyond the SM. Thus, in order 
to be compatible with the present neutrino data we choose here to go beyond the SM using one of its simplest and more appealing extensions,
the inverse seesaw model (ISS)~\cite{Mohapatra:1986aw,Mohapatra:1986bd,Bernabeu:1987gr}. This ISS extends the SM particle content by adding pairs of right-handed (RH) neutrinos
with opposite lepton number whose masses and couplings can be properly chosen to produce the physical light neutrino masses and oscillations in good agreement
with present data~\cite{Tortola:2012te,GonzalezGarcia:2012sz,Capozzi:2013csa}. In contrast to the original
seesaw type I model~\cite{Minkowski:1977sc,GellMann:1980vs,Yanagida:1979as,Mohapatra:1979ia,Schechter:1980gr}, the seesaw mechanism that produces
the small light physical neutrino masses in the ISS is associated to the smallness of the Majorana mass model parameters, such that when these are set to zero
lepton number conservation is restored, therefore increasing the symmetries of the model.  Another appealing feature of the ISS is that 
it allows for large Yukawa neutrino couplings while having at the same time moderately heavy right-handed neutrino masses at the ${\cal O}({\rm TeV})$ energies that are 
reachable at the present colliders, like the LHC. In addition to the possibility of being directly produced at colliders, these right-handed neutrinos could also lead to
a new rich phenomenology in connection with lepton flavor violating (LFV) Physics. This is because the ISS right-handed neutrinos can produce non-negligible contributions
to LFV processes via radiative corrections that are mediated by the sizable neutrino Yukawa couplings, therefore leading to clear signals/imprints in these rare processes,
which are totally absent in the SM . These LFV processes include the most frequently studied radiative decays,
$\mu \to e \gamma$, $\tau \to \mu \gamma$, $\tau \to e \gamma$, others like $\mu \to 3 e$, leptonic and semileptonic $\tau$ decays,
$\mu-e$ conversion in heavy nuclei, and others (see~\cite{Bernstein:2013hba} for a review). The ISS mechanism also has implications on deviations
from lepton flavor universality~\cite{Abada:2012mc, Abada:2013aba} and from lepton number conservation~\cite{Blennow:2010th,Chen:2011hc,LopezPavon:2012zg,Awasthi:2013we,Abada:2014vea}. 
Although quite promising future sensitivities for some of these LFV processes are expected, for instance, for $\mu-e$ conversion in heavy
nuclei~\cite{Abrams:2012er,Aoki:2012zza,Kuno:2013mha,PrismLOI20}, at present the highest sensitivity to LFV signals is obtained in $\mu \to e \gamma$ where
MEG has set an upper bound at branching ratio (BR), BR$(\mu \to e \gamma) < 5.7 \times 10^{-13}$~\cite{Adam:2013mnn}.   

In this paper, we study other LFV processes, the Higgs decays into lepton-antilepton pairs $H \to l_k\bar l_m$ with $k \neq m$, which are of obvious interest at present,
given the recent discovery of the Higgs particle and the fact that these rare Higgs decays are also being  presently explored at the LHC.
The current direct search at the LHC for these LFV Higgs decays (LFVHD) has been recently reported in \cite{CMS:2014hha}, where an upper limit of BR$(H\to\mu\tau)<1.57\%$ at $95\%$ C.L. has been set using $19.7 ~\mathrm{fb}^{-1}$ of $\sqrt s=8$ TeV.
This improves previous constraints from indirect measurements at LHC \cite{Harnik:2012pb} by roughly one order of magnitude (see also \cite{Blankenburg:2012ex}), and it is close to the previous estimates in \cite{Davidson:2012ds} that predicted sensitivities  of $4.5 \times 10^{-3}$ (see also, \cite{Bressler:2014jta}).
The future perspectives for LFVHD searches are encouraging due to the expected high statistics of Higgs events at future hadronic and leptonic colliders. 
Although, to our knowledge, there is no realistic study, including background estimates, of the expected future experimental sensitivities for these kinds of rare LFVHD events, a naive extrapolation from the present situation can be done.  
For instance, the future LHC runs with $\sqrt s=14$~TeV and total integrated luminosity of first $300~\rm{fb}^{-1}$ and later $3000~\rm{fb}^{-1}$ expect the production of about 25 and 250  million Higgs events, respectively, to be compared with 1 million Higgs events that the LHC produced after the first run \cite{futureLHCruns}. These large numbers suggest an improvement in the  long-term sensitivities to BR$(H\to\mu\tau)$ of at least two orders of magnitude with respect to the present sensitivity.
Similarly, at the planned lepton colliders, like the international linear collider (ILC) with\footnote{We thank J. Fuster for private communication with the updated ILC perspectives.} $\sqrt{s}=1$ TeV and $2.5~\rm{ab}^{-1}$\cite{Baer:2013cma}, and the future electron-positron circular collider (FCC-ee) as the TLEP with $\sqrt{s}=350$ GeV and $10~\rm{ab}^{-1}$\cite{Gomez-Ceballos:2013zzn},
the expectations are of about 1 and 2 million Higgs events, respectively,  with much lower backgrounds due to the cleaner environment, which will also allow for a large improvement in LFV Higgs searches with respect to the current sensitivities.

We will present a full one-loop
computation of the LFV partial decay widths, $\Gamma (H \to l_k\bar l_m)$, within the ISS context with three extra pairs of right-handed neutrinos, and will analyze in
full detail the predictions for the LFVHD rates,  BR$(H \to l_k\bar l_m)$,  as functions of the various relevant ISS parameters. These LFV Higgs decays
were analyzed in the context of the SM enlarged with three heavy Majorana neutrinos for the first time in~\cite{Pilaftsis:1992st}. Later, they were computed in the context
of the seesaw I model in~\cite{Arganda:2004bz}, and they were found to lead to extremely small rates due to the strong suppression from the extremely heavy right-handed 
neutrino masses, at $10^{14-15}$ GeV, in that case. This motivates our study of the LFV Higgs decays in the ISS case with the right-handed neutrino masses lying in contrast
at the ${\cal O}({\rm TeV})$ energy scale and therefore the rates are expected to be larger than in the seesaw I case.  The interest of neutrino masses at
this $\mathcal O(\rm{TeV})$ energy scale is also because they can be directly produced at the LHC.
Furthermore, we will also study in parallel the correlated one-loop predictions for the radiative decays, BR($l_m \to l_k \gamma$), 
within this same ISS context, and we will require full compatibility of our predictions with the present experimental upper bounds for the three relevant radiative decays,
$\mu \to e \gamma$,  $\tau \to \mu \gamma$, and $\tau \to e \gamma$, the first one being the most constraining one. 
We will require in addition that the input parameters of the ISS are compatible with the present neutrino data and other constraints, like perturbativity of the
neutrino Yukawa couplings.   After exploring the ISS parameter space we will conclude on the maximum allowed LFV Higgs decay rates within the ISS.

The paper is organized as follows: in section \ref{th-framework} we summarize our theoretical framework and shortly review the main features of the ISS that are relevant
for the present computation.  In section \ref{Widths} we present our computation of the one-loop LFV Higgs decay widths within the ISS and include, for completeness and
comparison, both the analytical formulas for the LFV Higgs decays and the LFV radiative decays. The full one-loop analytical formulas for the LFV Higgs form factors are
collected in the Appendix. Section \ref{results} is devoted to the presentation of the numerical results of our computation and also includes the predictions for both kind of LFV processes,
the branching ratios for the LFV Higgs decays, $H\to\mu \bar \tau$, $H\to e \bar \tau$, and $H\to e \bar \mu$ that we compare with the branching ratios for the radiative decays,
$\tau \to \mu \gamma$, $\tau \to e \gamma$,
and $\mu \to e \gamma$. Finally, we summarize our conclusions in section \ref{conclusions}.
 
\section{Theoretical framework}
\label{th-framework}
One of the simplest extensions of the SM leading to nonzero neutrino masses and mixing is the addition of fermionic gauge singlets. As mentioned above, a very attractive
model is the ISS that supplements the SM with pairs of RH neutrinos, denoted here by $\nu_R$ and $X$, with opposite lepton number. While the minimal model that fits oscillation
data requires only two generations of RH neutrinos~\cite{Malinsky:2009df}, we consider here a more generic model containing three pairs of fermionic singlets. It extends
the SM Lagrangian with the following neutrino Yukawa interactions and mass terms:
\begin{equation}
 \label{LagrangianISS}
 \mathcal{L}_\mathrm{ISS} = - Y^{ij}_\nu \overline{L_{i}} \widetilde{\Phi} \nu_{Rj} - M_R^{ij} \overline{\nu_{Ri}^C} X_j - \frac{1}{2} \mu_{X}^{ij} \overline{X_{i}^C} X_{j} + h.c.\,,
\end{equation}
where $L$ is the SM lepton doublet, $\Phi$ is the SM Higgs doublet, $\widetilde{\Phi}=\imath \sigma_2 \Phi^*$, with $\sigma_2$ being the corresponding
Pauli matrix, $Y_\nu$ is the $3\times3$ neutrino Yukawa coupling matrix, $M_R$ is a lepton number conserving complex $3\times3$ mass matrix, and $\mu_X$ is
a Majorana complex $3\times3$ symmetric mass matrix that violates lepton number conservation by two units. Setting the latter to zero would restore the conservation of lepton number,
thus increasing the symmetry of the model. This makes the smallness of $\mu_X$ natural since it could be seen as the remnant of a symmetry broken
at a higher energy~\cite{'tHooft:1979bh}. Since a Majorana mass term of the type $\overline{\nu_{Ri}} \nu_{Rj}^C$ would only give subleading corrections
to the neutrino masses and the observables considered here, we have taken it to be zero, for simplicity.

After electroweak symmetry breaking, the $9\times 9$ neutrino mass matrix reads, in the electroweak interaction basis $(\nu_L^C\,,\;\nu_R\,,\;X)$,
\begin{equation}
\label{ISSmatrix}
 M_{\mathrm{ISS}}=\left(\begin{array}{c c c} 0 & m_D & 0 \\ m_D^T & 0 & M_R \\ 0 & M_R^T & \mu_X \end{array}\right)\,,
\end{equation}
with the $3\times3$ Dirac mass matrix given by $m_D=Y_\nu \langle \Phi\rangle$, and the Higgs vacuum expectation value is taken
to be $\langle \Phi\rangle=v = 174\,\mathrm{GeV}$. Since this mass matrix is complex and symmetric, it can be diagonalized using a $9\times 9$ unitary matrix $U_\nu$ according to
\begin{equation}
U^T_\nu M_{\mathrm{ISS}} U_\nu = \text{diag}(m_{n_1},\dots,m_{n_9})\,. \label{Unu}
\end{equation}
This gives three light mass eigenstates and six heavy mass eigenstates, and the electroweak eigenstates and the mass eigenstates are related through
\begin{equation}
  \left(\begin{array}{c} \nu_L^C \\ \nu_R \\ X \end{array} \right) = U_\nu P_R \left(\begin{array}{c} n_1 \\ \vdots \\ n_9 \end{array} \right)\,\quad,\quad
    \left(\begin{array}{c} \nu_L \\ \nu_R^C \\ X^C \end{array} \right) = U_\nu^* P_L \left(\begin{array}{c} n_1 \\ \vdots \\ n_9 \end{array} \right)\,.
\end{equation}
In order to illustrate more simply the dependence on the seesaw parameters, let us first consider the one generation case and then we will come back to the three generation case.
In this one generation case there are just three ISS model parameters, $M_R$, $\mu_X$, and $Y_\nu$, and there are just three physical eigenstates: one light $\nu$ and
two heavy $N_1$ and $N_2$. In the limit $\mu_X \ll m_D, M_R$, the mass eigenvalues are given by:
\begin{align}
 m_\nu &= \frac{m_{D}^2}{m_{D}^2+M_{R}^2} \mu_X\,\label{mnu},\\
 m_{N_1,N_2} &= \pm \sqrt{M_{R}^2+m_{D}^2} + \frac{M_{R}^2 \mu_X}{2 (m_{D}^2+M_{R}^2)}\,,\label{mN}
\end{align}
with the light neutrino mass $m_\nu$ being proportional to $\mu_X$, thus making it naturally small, and the two heavy masses $m_{N_1,N_2}$ being close to each other.
As a consequence in this $\mu_X \ll m_D, M_R$ limit, these two nearly degenerate heavy neutrinos combine to form pseudo-Dirac fermions.

A similar pattern of neutrino mass eigenvalues occurs in the three generation case, with one light and two nearly degenerate heavy neutrinos per generation.
This can be illustrated clearly in the limit $\mu_X \ll m_D \ll M_R$, where the mass matrix $M_{\mathrm{ISS}}$ can be diagonalized by blocks~\cite{GonzalezGarcia:1988rw},
leading to the following $3 \times 3$ light neutrino mass matrix:
\begin{equation} \label{Mlight}
M_{\mathrm{light}} \simeq m_D {M_R^T}^{-1} \mu_X M_R^{-1} m_D^T\,,
\end{equation}
which is then diagonalized using the unitary Pontecorvo–-Maki-–Nakagawa–-Sakata (PMNS) matrix $U_{\rm PMNS}$\cite{PMNS}:
\begin{equation} \label{mnulight}
 U_{\rm PMNS}^T M_{\mathrm{light}} U_{\rm PMNS} = \mathrm{diag}(m_{\nu_1}\,, m_{\nu_2}\,, m_{\nu_3})\,,
\end{equation}
where $m_{\nu_1}$, $m_{\nu_2}$, and $m_{\nu_3}$ are the masses of the three lightest neutrinos.

Then, by defining a new $3 \times 3$ mass matrix by
\begin{equation}
M=M_R \mu_X^{-1} M_R^T,
\label{heavymasses} 
\end{equation}
the light neutrino mass matrix can be written similarly to the type I seesaw model as:
\begin{equation}
 M_{\mathrm{light}}\simeq m_D M^{-1} m_D^T\,.
\end{equation}
The mass pattern of the heavy neutrinos in the $\mu_X \ll m_D \ll M_R$ limit presents a similar behavior to the one generation case. The heavy neutrinos
form quasidegenerate pairs with a mass approximately given by the eigenvalues of $M_R$, namely $M_{R_{1,2,3}}$ for the first, second, and third generation, respectively,
and with a splitting of order $\mathcal O(\mu_X)$.

For our phenomenological purposes, and in order to implement easily the compatibility with present neutrino data, we will use here the useful
Casas-Ibarra parametrization~\cite{Casas:2001sr} that can be directly applied to the inverse seesaw model case, giving
\begin{equation}
 m_D^T= V^\dagger \mathrm{diag}(\sqrt{M_1}\,,\sqrt{M_2}\,,\sqrt{M_3})\; R\; \mathrm{diag}(\sqrt{m_{\nu_1}}\,, \sqrt{m_{\nu_2}}\,, \sqrt{m_{\nu_3}}) U^\dagger_{\rm PMNS}\,,
\label{CasasIbarraISS}
\end{equation}
where $V$ is a unitary matrix that diagonalizes $M$ according to $ M=V^\dagger \mathrm{diag}(M_1\,, M_2\,, M_3) V^*$ and $R$ is a complex orthogonal matrix that can be written as
\begin{equation}
\label{R_Casas}
R = \left( \begin{array}{ccc} c_{2} c_{3} 
& -c_{1} s_{3}-s_1 s_2 c_3& s_{1} s_3- c_1 s_2 c_3\\ c_{2} s_{3} & c_{1} c_{3}-s_{1}s_{2}s_{3} & -s_{1}c_{3}-c_1 s_2 s_3 \\ s_{2}  & s_{1} c_{2} & c_{1}c_{2}\end{array} \right) \,,
\end{equation}
where $c_i\equiv \cos \theta_i$, $s_i\equiv \sin\theta_i$ and
$\theta_1$, $\theta_2$, and $\theta_3$ are arbitrary complex angles. 

In summary, assuming $\mu_X={\rm diag}(\mu_{X_1}, \mu_{X_2}, \mu_{X_3})$ and $M_R={\rm diag}(M_{R_1},M_{R_2},M_{R_3})$ (hence, diagonal $M$),
the input ISS parameters that will have to be fixed for our forthcoming study of the LFV rates are the 
following: $m_{\nu_{1,2,3}}$,  $\mu_{X_{1,2,3}}$, $M_{R_{1,2,3}}$, $\theta_{1,2,3}$, and the entries of the $U_{\rm PMNS}$ matrix.  
For all the numerical analysis in this work, and in order to keep agreement with the experimental neutrino data, we will choose the lightest neutrino mass,
here assumed to be $m_{\nu_1}$, as a free input parameter and the other two light masses will be obtained from the two experimentally measured mass differences:
\begin{equation}
\label{mnu2mnu3}
m_{\nu_2}=\sqrt{m_{\nu_1}^2+\Delta m_{21}^2}\ ,\quad 
m_{\nu_3}=\sqrt{m_{\nu_1}^2+\Delta m_{31}^2}\ .
\end{equation}

Similarly, the three light neutrino mixing angles will also be set to their measured values. For simplicity, we will set to zero the CP-violating phase of the $U_{\rm PMNS}$ matrix.
Specifically, we have used the results of the global fit~\cite{GonzalezGarcia:2012sz} leading to 
\begin{align}
 \sin^2\theta_{12}&=0.306^{+0.012}_{-0.012}\,, 					 & \Delta m^2_{21}&=7.45^{+0.19}_{-0.16}\times10^{-5}\;\mathrm{eV}^2\,,\nonumber\\
 \sin^2\theta_{23}&=0.446^{+0.008}_{-0.008}\,, 					&  \Delta m^2_{31}&=2.417^{+0.014}_{-0.014}\times10^{-3}\;\mathrm{eV}^2\,,\\
\sin^2\theta_{13}&=0.0231^{+0.0019}_{-0.0019}\,,				&\nonumber
\end{align}
where we have assumed a normal hierarchy. Regarding the input lightest neutrino mass, $m_{\nu_1}$, we have chosen it so that the effective electron neutrino mass in
$\beta$ decay agrees with the upper limit from the Mainz and Troitsk experiments~\cite{Kraus:2004zw, Aseev:2011dq},
\begin{equation}
 m_\beta < 2.05\;\mathrm{eV}\quad\mathrm{at}\;95\%\;\mathrm{C.L.}
\end{equation}
For the final numerical evaluation of the eigenvalues and eigenstates of the full $9 \times 9$ neutrino matrix, we have used our private {\tt Mathematica} code
that solves this system numerically, using all the previously mentioned input parameters and experimental data; and besides it also computes the Yukawa coupling matrix
entries by using eq.~(\ref{CasasIbarraISS}). 

In order to illustrate the kind of generic neutrino spectra that one obtains in the ISS and that indeed follow the previously commented pattern, we have chosen in this
section to show three examples of spectra whose most relevant parameters for the present work are collected in table~\ref{spectra}.  

\begin{table}[ht]
\begin{center}
\begin{tabular}{|c|c|c|c|}
\hline
ISS examples & A & B & C\\
\hline
$M_{R_1}({\rm GeV})$ & $1.5\times10^{4}$ & $ 1.5\times10^{2} $ & $ 1.5\times10^{2} $\\
$M_{R_2}({\rm GeV})$ & $1.5\times10^{4}$ & $ 1.5\times10^{3} $ & $ 1.5\times10^{3} $\\
$M_{R_3}({\rm GeV})$ & $1.5\times10^{4}$ & $ 1.5\times10^{4} $ & $ 1.5\times10^{4} $\\
$\mu_{X_{1,2,3}}({\rm GeV})$ & $5\times10^{-8} $ & $5\times10^{-8} $ & $5\times10^{-8} $\\
$m_{\nu_1}({\rm eV})$ & $0.1$ & $0.1$ & $0.1$\\
$\theta_{1,2,3}({\rm rad})$ & $0,0,0$ & $0,0,0$ & $ \pi/4,0,0$\\
\hline
$m_{n_1}({\rm eV})$ & $0.0998 $ & $0.0998 $ & $0.0998 $\\
$m_{n_2}({\rm eV})$ & $0.1002 $ & $0.1002 $ & $0.1002 $\\
$m_{n_3}({\rm eV})$ & $0.1112 $ & $0.1112 $ & $0.1112 $\\
$m_{n_4}({\rm GeV})$ & 15014.99250747 & 150.1499250500 & 150.1499250500\\
$m_{n_5}({\rm GeV})$ & 15014.99250752 & 150.1499250999 & 150.1499250999\\
$m_{n_6}({\rm GeV})$ & 15015.04822299 & 1501.504822277 & 1501.587676006\\
$m_{n_7}({\rm GeV})$ & 15015.04822304 & 1501.504822327 & 1501.587676056\\
$m_{n_8}({\rm GeV})$ & 15016.70543659 & 15016.70543659 & 15015.87685358\\
$m_{n_9}({\rm GeV})$ & 15016.70543664 & 15016.70543664 & 15015.87685363\\
\hline
$|(Y_\nu Y_\nu^\dagger)_{23}|$ & 0.8 & 8.0 & 1.4 \\
$|(Y_\nu Y_\nu^\dagger)_{12}|$ & 0.2 & 1.7 & 0.3 \\
$|(Y_\nu Y_\nu^\dagger)_{13}|$ & 0.2 & 1.8 & 4.0 \\
\hline
\end{tabular}
\caption{Examples of neutrino mass spectrum in the ISS for various input parameters. The relevant nondiagonal $|(Y_\nu Y_\nu^\dagger)_{km}|$ elements are also included.}\label{spectra}
\end{center}
\end{table}
We see clearly in these three examples that one typically gets the announced pattern of neutrino masses: three light neutrinos compatible with data and six heavy ones,
with their heavy masses being degenerate in pairs to values close to $M_{R_1}$, $M_{R_2}$, and $M_{R_3}$ respectively, and their tiny mass differences given approximately
by $\mu_{X_1}$, $\mu_{X_2}$, and $\mu_{X_3}$. We also see in this table that one can get sizable Yukawa couplings, in particular leading to large nondiagonal entries in flavor
space, which are the relevant ones for the present work on lepton flavor violation. It should also be noticed that the heavy masses that are governing the size of these off diagonal entries are not
those of $M_R$ but those of eq.~(\ref{heavymasses}), which are largely heavier, therefore leading in general to larger LFV rates in the ISS than in the seesaw I. For instance,
the chosen examples in this table lead to large $|(Y_\nu Y_\nu^\dagger)_{23}|$ and $|(Y_\nu Y_\nu^\dagger)_{13}|$ in the ${\cal O}(1-10)$ range. $|(Y_\nu Y_\nu^\dagger)_{12}|$ in
these examples is  slightly smaller, $\leq {\cal O}(1)$.

One way of checking the validity of the parametrization in eq.~(\ref{CasasIbarraISS}) is by comparing the input light neutrino mass values in this equation with the
lightest output mass values obtained as a solution of eq.~(\ref{Unu}).
We have checked that the error on the light neutrino masses estimated with this parametrization, meaning the differences between the input $m_{\nu_{1,2,3}}$ and
the output $m_{n_{1,2,3}}$ masses, is below $10\%$ and that the rotation matrix $U_\nu$ exhibits the required unitarity property. Furthermore, since a given set of input
parameters can generate arbitrarily large Yukawa couplings, we will enforce their perturbativity by setting an upper limit on the entries of the neutrino Yukawa coupling matrix,
given by
\begin{equation}\label{Ymax}
\frac{|Y_{ij}|^2}{4\pi}<1.5\,,
\end{equation}
for $i,j=1,2,3$.
  This particular perturbativity condition has been used in the literature (see, for instance, {\tt SPheno version 2.0}~\cite{Porod:2003um}) but others more conservative than this have also been used (see, for instance, {\tt SPheno version 3.1}~\cite{Porod:2011nf}).
In the absence of a concrete evaluation of the next order corrections to the observable of interest (two-loop contributions to the LFVHD rates in our present case, which are beyond the scope of this article), the perturbativity condition is not uniquely defined and the choice of a specific criterion is an open issue.
For instance, the use of a more conservative condition like $|Y_{ij}|^2 < 4\pi$ instead of eq.~(\ref{Ymax}) will not qualitatively change our results and will just lead to a decrease on the maximum LFVHD rates allowed by perturbativity\footnote{We thank the referee for suggesting this other possible choice.},  by roughly a factor of $(1/5)$, as can be easily estimated with our approximate formulas that will be presented later.

Finally, to complete our setup of the theoretical framework for our study of LFV, we also have to specify all the relevant interactions that will enter in the computation
of the LFVHD rates. We focus here on the relevant interactions involving neutrinos, which are the only ones that are assumed here to differ from those of the SM. These include
the neutrino Yukawa couplings, the gauge couplings of the charged gauge bosons $W^\pm$ to the lepton-neutrino pairs and the corresponding couplings of the charged Goldstone bosons,
denoted here by $G^{\pm}$,  to the lepton-neutrino pairs. In our one-loop computation of the LFV rates we will choose to work in the mass basis for all the particles involved,
with diagonal charged leptons, and taking into account the contributions from all the nine physical neutrinos. As for the gauge choice,  we will choose the Feynman-t'Hooft gauge.
Following the notation and presentation  in~\cite{Ilakovac:1994kj, Arganda:2004bz}, the relevant interactions are given in the mass basis by the 
following terms of the Lagrangian:
\begin{align}
\mathcal{L}_{\rm int}^{W^{\pm}} &= \frac{-g}{\sqrt{2}} W^{\mu -}\bar{l_i} B_{l_i n_j} \gamma_{\mu} P_L n_j + h.c\,,\nonumber\\ 
\mathcal{L}_{\rm int}^{H} &= \frac{-g}{2 m_W} H \bar n_i C_{n_i n_j} \left[ m_{n_i} P_L + m_{n_j} P_R \right] n_j\,,\nonumber\\ 
\mathcal{L}_{\rm int}^{G^{\pm}} &= \frac{-g}{\sqrt{2} m_W} G^{-}\left[\bar{l_i} B_{l_i n_j} (m_{l_i} P_L - m_{n_j} P_R) n_j \right] + h.c\,,
\end{align}
where $P_L$ and $P_R$ are respectively the left- and right-chirality projectors, given by $(1-\gamma^5)/2$ and $(1+\gamma^5)/2$, and the coupling factors
$B_{l_i n_j}$ ($i=1,\,2,\,3,\, j=1,\dots,9$) and $C_{n_i n_j}$ ($i,\,j=1,\dots,9$) are defined in terms of the $U_{\nu}$ matrix of eq.~(\ref{Unu}) by
\begin{eqnarray}
B_{l_i n_j} = U_{ij}^{\nu *}\,,\\
C_{n_i n_j} = \sum_{k=1}^3 U_{k i}^{\nu} U_{kj}^{\nu *}\ .
\end{eqnarray}

\section{Computation of the LFV decay widths}
\label{Widths}
\begin{figure}
\begin{center}
\includegraphics[width=\textwidth]{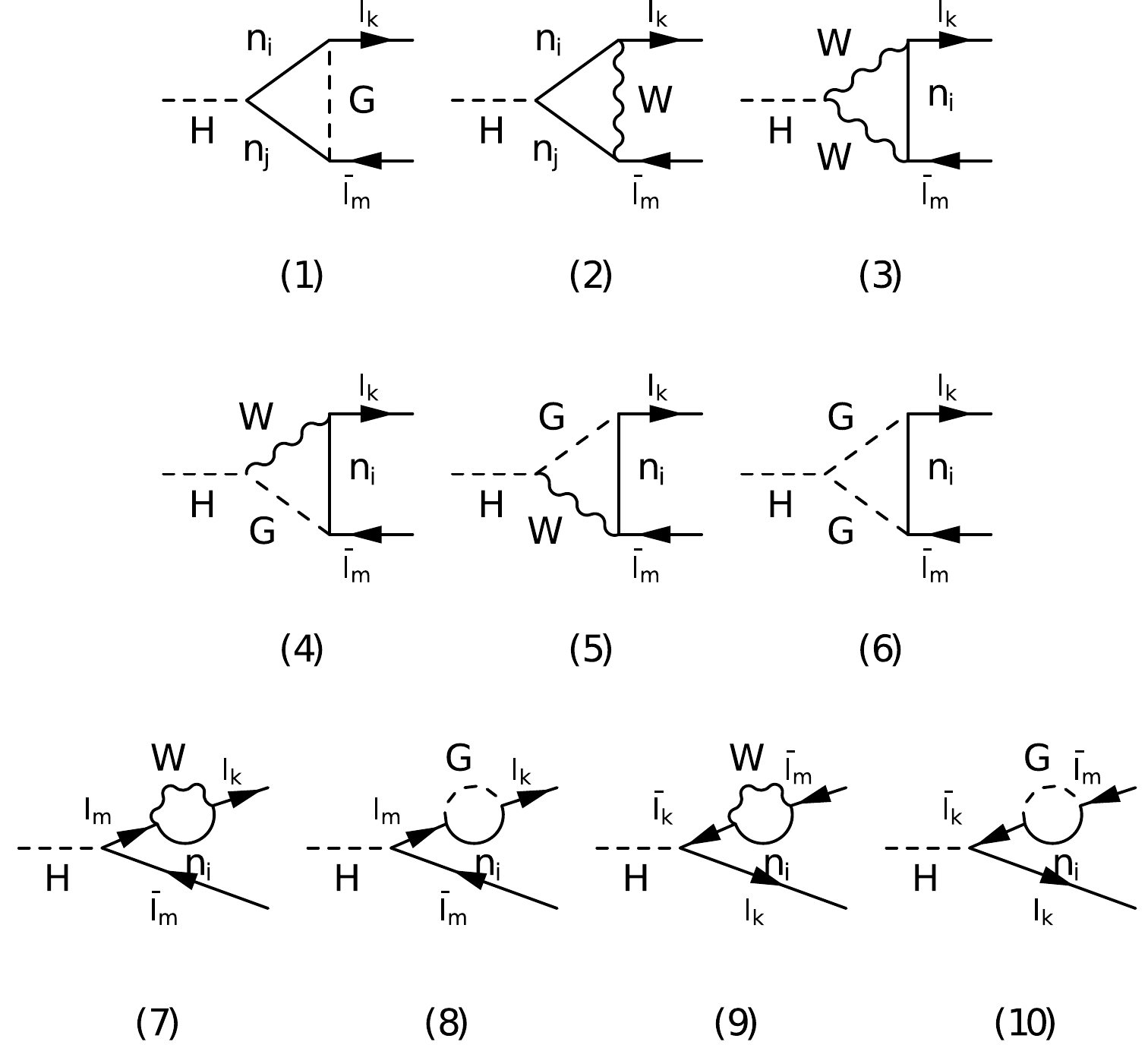}
\caption{One-loop contributing diagrams to the LFV Higgs decays 
$H \to l_k \bar l_m$ in the ISS with massive neutrinos $n_i\,(i=1,..,9)$.}
\label{diagrams}
\end{center}
\end{figure}

In the calculation of the LFV Higgs decay rates, we consider the full set of contributing one-loop diagrams, drawn in fig.~\ref{diagrams}, and adapt to our present ISS case
the complete one-loop formulas for the  $\Gamma(H \to l_k\bar l_m)$ partial decay width, taken from~\cite{Arganda:2004bz}, which we include, for completeness, also here.  The
relation between the form factors $F_L$ and $F_R$ given in the Appendix and the decay amplitude $F$ is given by
\begin{equation}
i F = -i g \bar{u}_{l_k} (-p_2) (F_L P_L + F_R P_R) v_{l_m}(p_3)\,,
\end{equation}
where
\begin{equation}
F_L = \sum_{i=1}^{10} F_L^{(i)},\,\,  F_R = \sum_{i=1}^{10} F_R^{(i)}\,,
\end{equation}
and $p_1=p_3-p_2$ is the ingoing Higgs boson momentum. 

The width for the LFV Higgs decays is obtained from these form factors by
\vspace{1cm}
\begin{eqnarray}
\Gamma (H \to {l_k} \bar{l}_m)& = &\frac{g^2}{16 \pi m_H} 
\sqrt{\left(1-\left(\frac{m_{l_k}+m_{l_m}}{m_H}\right)^2\right)
\left(1-\left(\frac{m_{l_k}-m_{l_m}}{m_H}\right)^2\right)} \nonumber\\
&&\times\left((m_H^2-m_{l_k}^2-m_{l_m}^2)(|F_L|^2+|F_R|^2)- 4 m_{l_k} m_{l_m} Re(F_L F_R^*)\right)\,.
\label{decay}
\end{eqnarray}
In this work we focus on the decays $H\to \mu\bar\tau,e\bar\tau,e\bar\mu$ and do not consider their related $CP$ conjugate decays $H\to\tau\bar\mu,\tau\bar e,\mu \bar e$, which,
in the presence of complex phases, could lead to different rates.

We have explicitly checked that the only divergent contributions to the LFV Higgs decays arise from the diagrams (1), (8), and (10), and that they cancel among each other,
in agreement with~\cite{Arganda:2004bz}, giving rise to a total finite result. All these formulas for the LFV Higgs form factors and the LFV Higgs partial decay widths have been
implemented into our private {\tt Mathematica} code. In order to get numerical predictions for the BR$(H \to {l_k} \bar{l}_m)$ rates we use
$m_H=126\,{\rm GeV}$ and its corresponding SM total width is computed with {\tt FeynHiggs}~\cite{Heinemeyer:1998yj,Heinemeyer:1998np,Degrassi:2002fi} including two-loop corrections.  

At the same time that we analyze the LFV Higgs decays, we also compute the one-loop $l_m \to l_k \gamma$ decay rates within this same ISS framework and for the same input parameters,
and check that these radiative decay rates are compatible with their present experimental $90\%$ C.L. upper  bounds:
\begin{align}
{\rm BR}(\mu\to e\gamma)&\leq 5.7\times 10^{-13}\text{\cite{Adam:2013mnn}}\label{MUEGmax}\,,\\
{\rm BR}(\tau\to e\gamma)&\leq 3.3\times 10^{-8}~\text{\cite{Aubert:2009ag}}\label{TAUEGmax}\,,\\
{\rm BR}(\tau\to \mu\gamma)&\leq 4.4\times 10^{-8}~\text{\cite{Aubert:2009ag}}\label{TAUMUGmax}\,.
\end{align}
In order to calculate these LFV radiative decay rates, which have been first computed in~\cite{GonzalezGarcia:1991be}, we use the analytical formulas
appearing in~\cite{Ilakovac:1994kj} and~\cite{Deppisch:2004fa} that have also been implemented in our code:
\begin{equation}
\label{BRradiative}
{\rm BR}(l_m\to l_k\gamma)=\frac{\alpha^3_W s_W^2}{256\pi^2}\left(\frac{m_{l_m}}{M_W}\right)^4\frac{m_{l_m}}{\Gamma_{l_m}}|G_{mk}|^2,
\end{equation}
where $\Gamma_{l_m}$ is total decay width of the lepton $l_m$, and
\begin{align}\label{Gmk}
G_{mk}&=\sum_{i=1}^{6}B^*_{mi}B_{ki}\, G_\gamma\left(\dfrac{m_{N_i}^2}{M_W^2}\right)\,,\nonumber\\
G_\gamma(x)&=-\dfrac{2x^3+5x^2-x}{4(1-x)^3}-\dfrac{3x^3}{2(1-x)^4}\log x\,,
\end{align} 
where the sum above extends over the six heavy neutrinos, $N_{1,..,6}=n_{4,..,9}$.  Notice that in the above formulas (\ref{BRradiative})-(\ref{Gmk}) the mass of the
final lepton $l_k$ has been neglected. 

Finally, we offer a few words summarizing the various constraints that we have also implemented in our code.
As we have already said, we have imposed the perturbativity constraint on the neutrino Yukawa couplings given in eq.~(\ref{Ymax}). Regarding the Higgs total width, it could be
modified by
the presence of sterile neutrinos with a mass below the Higgs boson mass that could open new invisible decays, as was studied in~\cite{BhupalDev:2012zg,Cely:2012bz}. However,
in this work, we focus on the scenario where the new fermionic singlets have a mass above $200\,\mathrm{GeV}$, thus escaping these constraints. If the right-handed neutrinos
provide a sizable contribution to LFV processes, a non-negligible contribution to the lepton electric dipole moments (EDMs) could also be expected in the general case with complex
phases. Thus, to avoid potential constraints from EDMs, we assume in most of this work that all mass matrices are real, as well as the PMNS matrix. The case of complex $R$ matrix
has also been considered in this work, but as it will be shown later (see fig.~\ref{BRs_theta1_degenerate}) it is highly constrained by $\mu\to e\gamma$.
Additional constraints might also arise from lepton universality tests. However, in the scenario that we consider where the sterile neutrinos are heavier than the Higgs boson,
points that would be excluded by lepton universality tests are already excluded by $\mu \rightarrow e \gamma$, as can be seen in fig.~8 of~\cite{Abada:2013aba}. 
In the end, we found that the most constraining observable for our study is by far $\mu \rightarrow e \gamma$.

\section{Numerical results for the LFV rates}
\label{results}

In this section we present our numerical results for the LFV Higgs decay rates, 
BR$(H \to \mu \bar \tau)$, BR$(H \to e \bar \tau)$, and BR$(H \to e \bar \mu)$, and we also compare them with the numerical results for the related radiative decay rates,
BR$(\mu \to e \gamma)$, 
BR$(\tau \to e \gamma)$, and BR$(\tau \to \mu \gamma)$. First, we consider the simplest case of diagonal $M_R$ and $\mu_X$ matrices and study all these LFV rates 
as functions of the more relevant ISS parameters, namely, $M_{R_i}$, $\mu_{X_i}$, $m_{\nu_i}$, and the $R$ matrix angles, $\theta_i$, trying to localize the areas of
the parameter space where the LFV Higgs decays can be both large and respect the constraints on the radiative decays.  
The results of this first case will be presented in two generically different scenarios for the heavy neutrinos: (1) the case of (nearly) degenerate heavy neutrinos
(first subsection), and (2) the case of hierarchical heavy neutrinos (second subsection).  In the last subsection, we then consider the most general case of nondiagonal
$\mu_X$ and look for solutions within the ISS that lead to the largest and allowed LFVHD rates. We will then present our predictions for the maximal allowed
BR$(H \to \mu \bar \tau)$ and 
BR$(H \to e \bar \tau)$ rates and will provide some specific examples for this kind of ISS scenarios. 
\subsection{Degenerate heavy neutrinos}
\label{degenerate}

\begin{figure}[t!]
\begin{center}
\begin{tabular}{cc}
\includegraphics[width=0.49\textwidth]{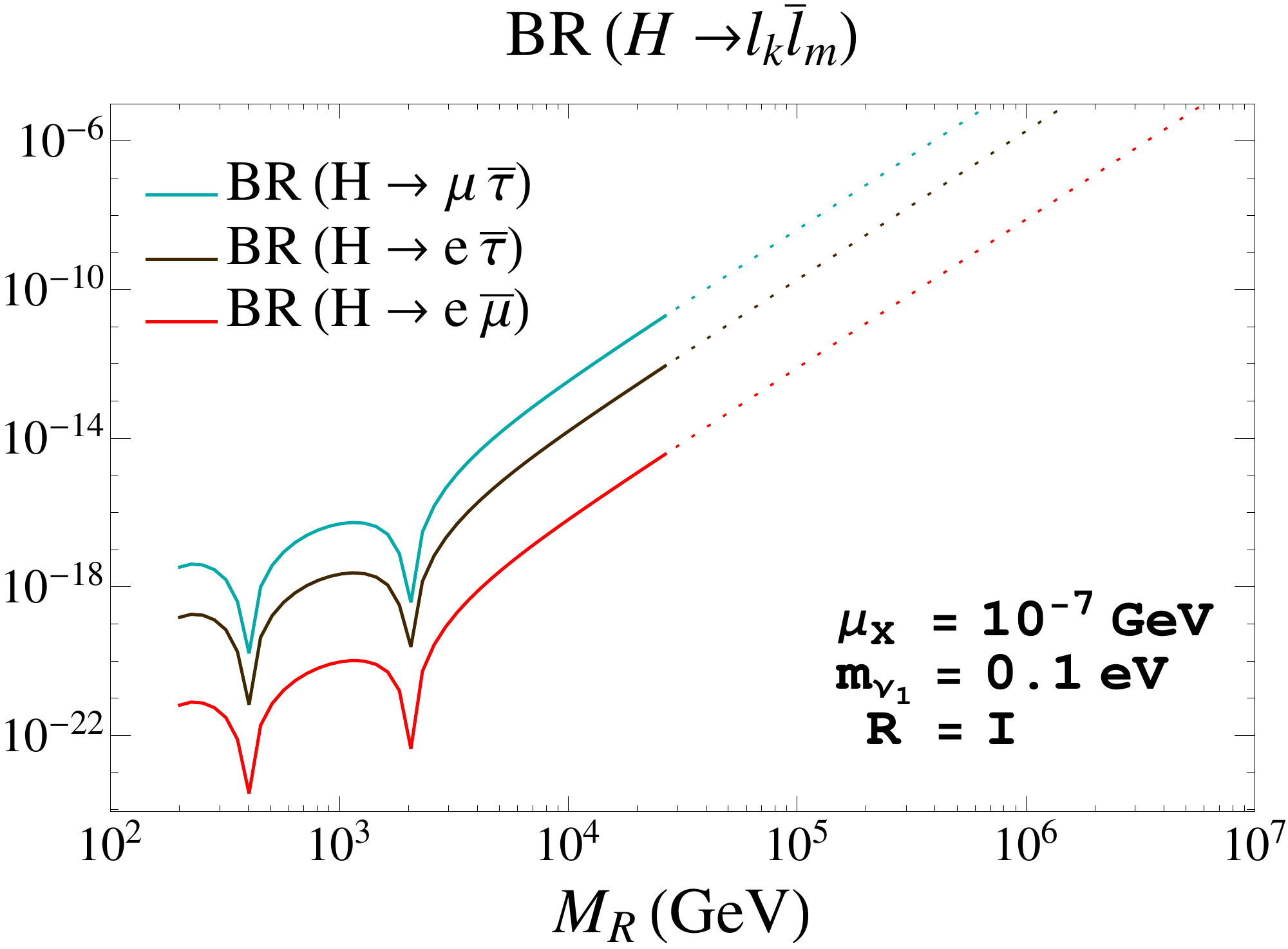} &
\includegraphics[width=0.49\textwidth]{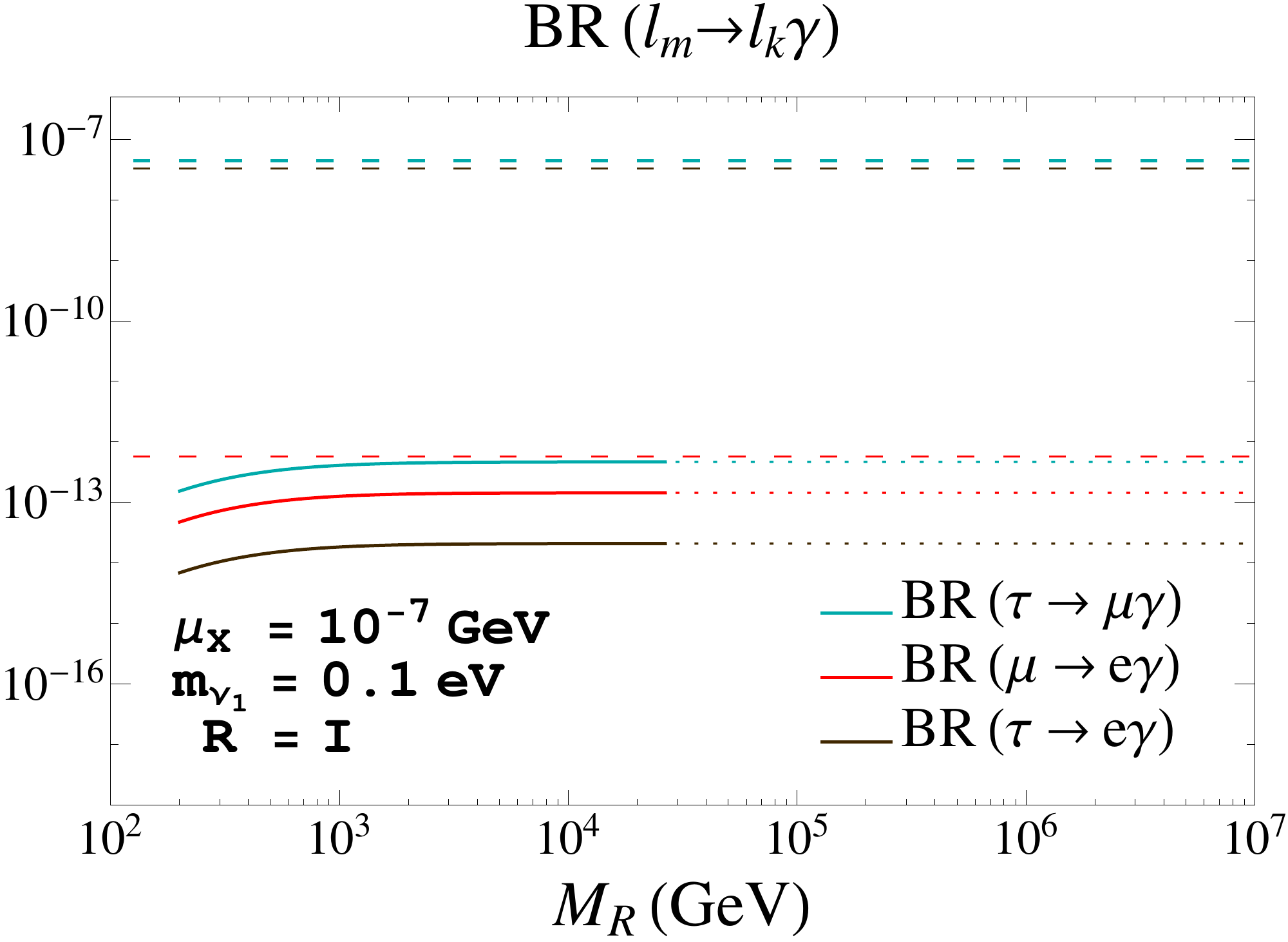}
\end{tabular}
\caption{Predictions for the LFV decay rates as functions of $M_R$ in the  degenerate heavy neutrinos case. Left panel: BR$(H \to \mu \bar \tau)$ (upper blue line),
BR$(H \to e \bar \tau)$ (middle dark brown line), BR$(H \to e \bar \mu)$ (lower red line). Right panel: BR$(\tau \to \mu \gamma)$ (upper blue line), BR$(\mu \to e \gamma)$
(middle red line), BR$(\tau \to e \gamma)$ (lower dark brown line). The other input parameters are set to $\mu_X= 10^{-7} \, {\rm GeV}$, 
$m_{\nu_1}=0.1 \, {\rm eV}$, $R=I$. The dotted lines in both panels indicate nonperturbative neutrino Yukawa couplings. The horizontal dashed lines in the right panel
are the present ($90\%$ C.L.) upper bounds on the radiative decays: BR$(\tau \to \mu \gamma)<4.4 \times 10^{-8}$~\cite{Aubert:2009ag} (blue line), 
BR$(\tau \to e \gamma)<3.3 \times 10^{-8}$~\cite{Aubert:2009ag} (dark brown line), BR$(\mu \to e \gamma)< 5.7 \times 10^{-13}$~\cite{Adam:2013mnn} (red line).
}\label{ALL_LFVdecays}
\end{center}
\end{figure}

The case of (nearly) degenerate heavy neutrinos is implemented here by choosing 
degenerate entries in $M_R={\rm diag}(M_{R_1},M_{R_2},M_{R_3})$ and in $\mu_X={\rm diag}(\mu_{X_1},
\mu_{X_2},\mu_{X_3})$, i.e., by setting $M_{R_i} = M_R$ and $\mu_{X_i} = \mu_X$ ($i = 1, 2, 3$). 

First we show in fig.~\ref{ALL_LFVdecays} the results for all the LFV rates as functions of the common right-handed neutrino mass parameter $M_R$ for all
the LFV Higgs decay channels (left panel) and for all the LFV radiative decay channels (right panel). Here we have fixed the other input parameters
to $\mu_X=10^{-7}$ GeV, $m_{\nu_1}=0.1$ eV, and $R=I$. As expected, we find that the largest LFV Higgs decay rates are for BR($H \to \mu\bar \tau$) and the
largest radiative decay rates are for BR($\tau \to \mu \gamma$). We also see that, for this particular choice of input parameters, all the predictions for the
LFV Higgs decays are allowed by the present experimental upper bounds on the three radiative decays (dashed horizontal lines in all our plots for the radiative decays)
for all explored values of $M_R$ in the interval $(200, 10^7)\, {\rm GeV}$. Besides, it shows clearly that the most constraining radiative decay at present is by far
the $\mu \to e \gamma$ radiative decay. This is so in all the cases 
explored in this work, so whenever we wish to conclude on the allowed LFVHD rates we will focus mainly on this radiative channel. 

Regarding the $M_R$ dependence shown in fig.~\ref{ALL_LFVdecays}, we clearly see that the LFVHD rates grow faster with $M_R$ than the radiative decays which indeed tend
to a constant value for $M_R$ above $\sim 10^3$ GeV. In fact, the LFVHD rates can reach quite sizable values at the large $M_R$ region of these plots, yet are allowed by the
constraints on the radiative decays. For instance, we obtain BR$(H \to \mu\bar \tau) \sim 10^{-6}$ for $M_R = 4\times 10^5$ GeV. However, our requirement of perturbativity
for the neutrino Yukawa coupling entries, see eq.~(\ref{Ymax}), does not allow for such large $M_R$ values leading to too-large $Y_\nu$ values in the framework of our parametrization
of eq.~(\ref{CasasIbarraISS}). Indeed, the exclusion region for $M_R$ from perturbativity of $Y_\nu$ (given by the dotted lines in these plots), forbids these large $M_R$ values.
For the specific input parameter values of this fig.~\ref{ALL_LFVdecays}, the forbidden values are for $M_R$ above $3 \times 10^4$ GeV, and this leads to maximum 
allowed values of BR$(H \to \mu\bar \tau) \sim 2 \times10^{-11}$, BR$(H \to e \bar \tau) \sim 10^{-12}$, and BR$(H \to e \bar \mu) \sim 5 \times 10^{-15}$.          

\begin{figure}[t!]
\begin{center}
\begin{tabular}{cc}
\includegraphics[width=0.49\textwidth]{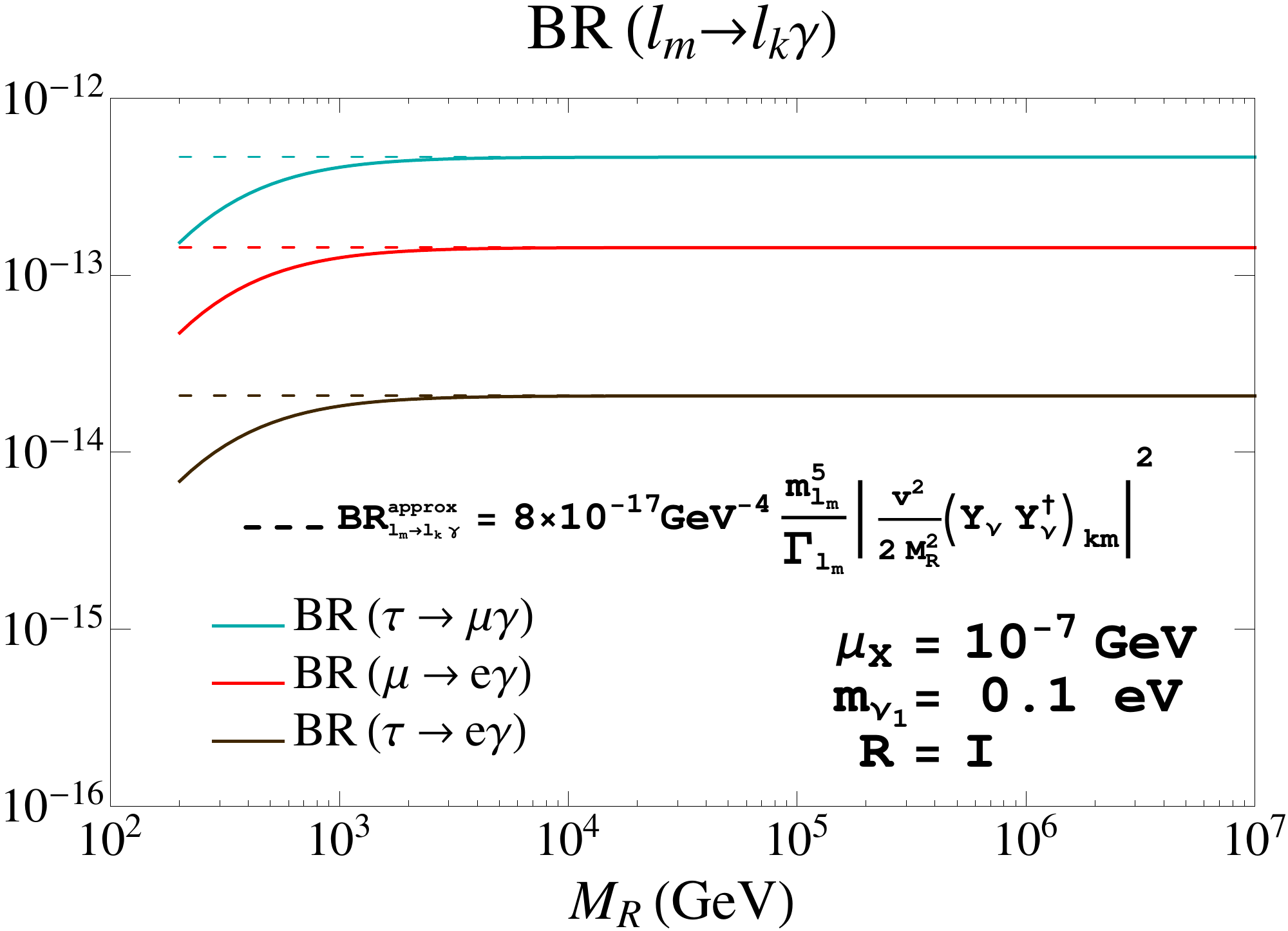} &
\includegraphics[width=0.49\textwidth]{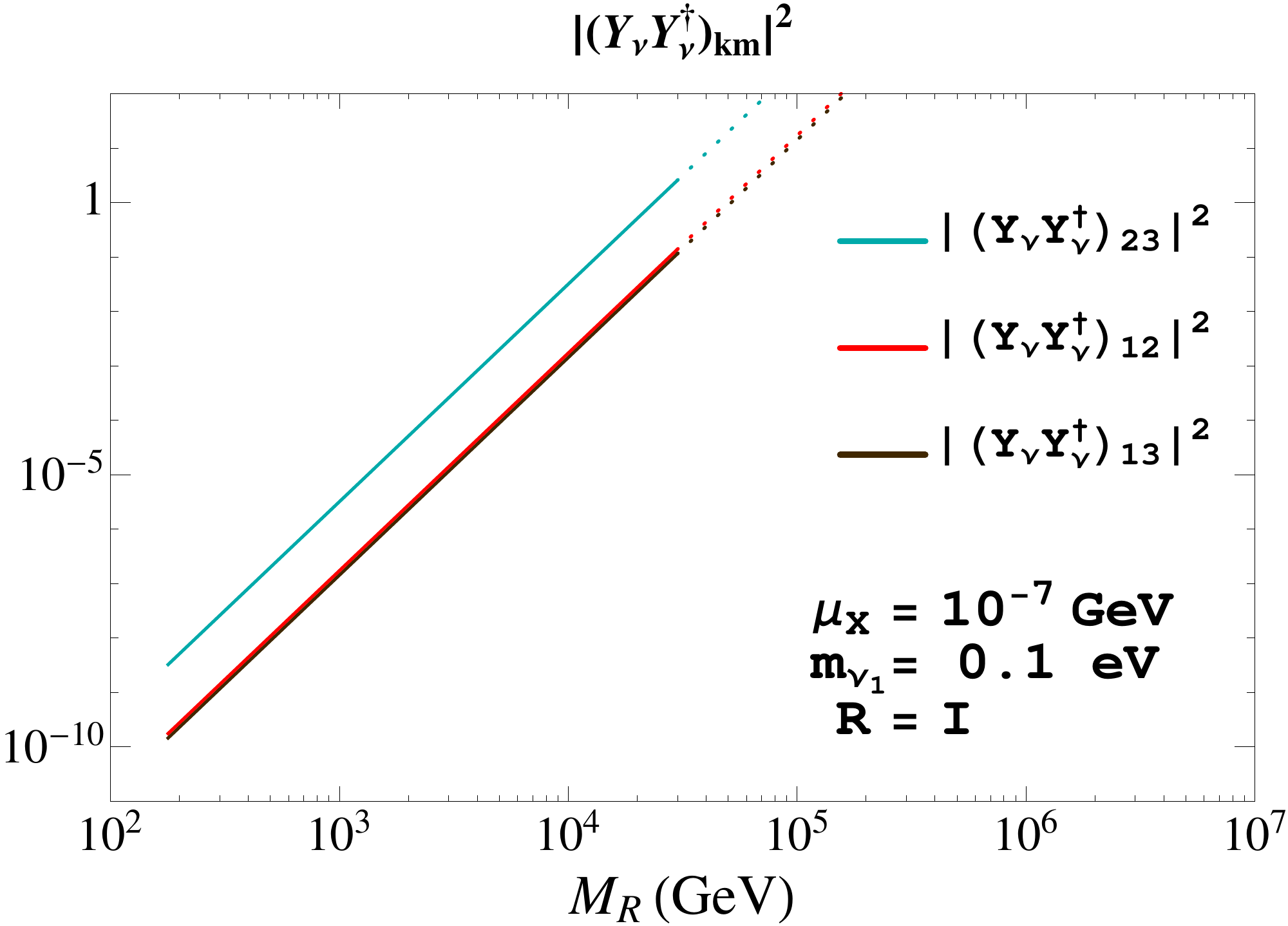}
\end{tabular}
\caption{Comparison of the full one-loop and approximate rates for the radiative decays $l_m \to l_k \gamma$ and their relation with the $(Y_\nu Y_\nu^\dagger)_{km}$ nondiagonal
matrix elements in the degenerate heavy neutrinos case. Left panel: full one-loop rates (solid lines) and approximate rates (dashed lines) as functions of $M_R$. Right panel:
$|(Y_\nu Y_\nu^\dagger)_{km}|^2$ versus $M_R$ for $km=23$ (blue line), $km=12$ (red line), and $km=13$ (dark brown line). Dotted lines indicate nonperturbative neutrino Yukawa
couplings. The other input parameters are set to $\mu_X= 10^{-7} \, {\rm GeV}$, 
$m_{\nu_1}=0.1 \, {\rm eV}$, and $R=I$.}\label{approxrad}
\end{center}
\end{figure}

The qualitatively different functional behavior with $M_R$ of the LFVHD and the radiative rates shown by fig.~\ref{ALL_LFVdecays} is an interesting feature
that we wish to explore further.  Whereas the BR$(l_m\to~l_k\gamma)$ rates follow the expected behavior with $M_R$ as derived from their dependence with the relevant
$(Y_\nu Y_\nu^\dagger)_{km}$ element, the  
BR($H \to l_k \bar l_m$) rates do not follow this same pattern. As it is clearly illustrated in 
fig.~\ref{approxrad}, the radiative decay rates can be well approximated for large $M_R$ by a simple  function of $|(Y_\nu Y_\nu^\dagger)_{km}|^2$ given by
\begin{equation}
{\rm BR}^{\rm approx}_{l_m \to l_k \gamma}=8 \times 10^{-17} \frac{m_{l_m}^5({\rm GeV}^5)}{\Gamma_{l_m}{\rm (GeV)}} 
\bigg|\frac{v^2}{2M_R^2}(Y_\nu Y_\nu^\dagger)_{km}\bigg|^2,
\label{approxformula} 
\end{equation} 
which provides predictions very close to the exact rates (given by the solid lines) for $M_R>10^3$ GeV. Then we can understand the final constant behavior of all the radiative
decay rates with $M_R$, since the $|(Y_\nu Y_\nu^\dagger)_{km}|^2$ elements grow with $M_R$ approximately as $M_R^4$  in the parametrization here used of eq.~(\ref{CasasIbarraISS}),
as can be seen in the plot on the right in fig.~\ref{approxrad}.  
This simple behavior with $M_R$ is certainly not the case of the LFVHD rates, and we conclude that these do not follow this same behavior with $|(Y_\nu Y_\nu^\dagger)_{km}|^2$.
This different functional behavior of 
BR$(H \to l_k \bar l_m)$ with $M_R$ will be further explored and clarified later.

\begin{figure}[t!] 
\begin{center}
\begin{tabular}{cc}
\includegraphics[width=0.49\textwidth]{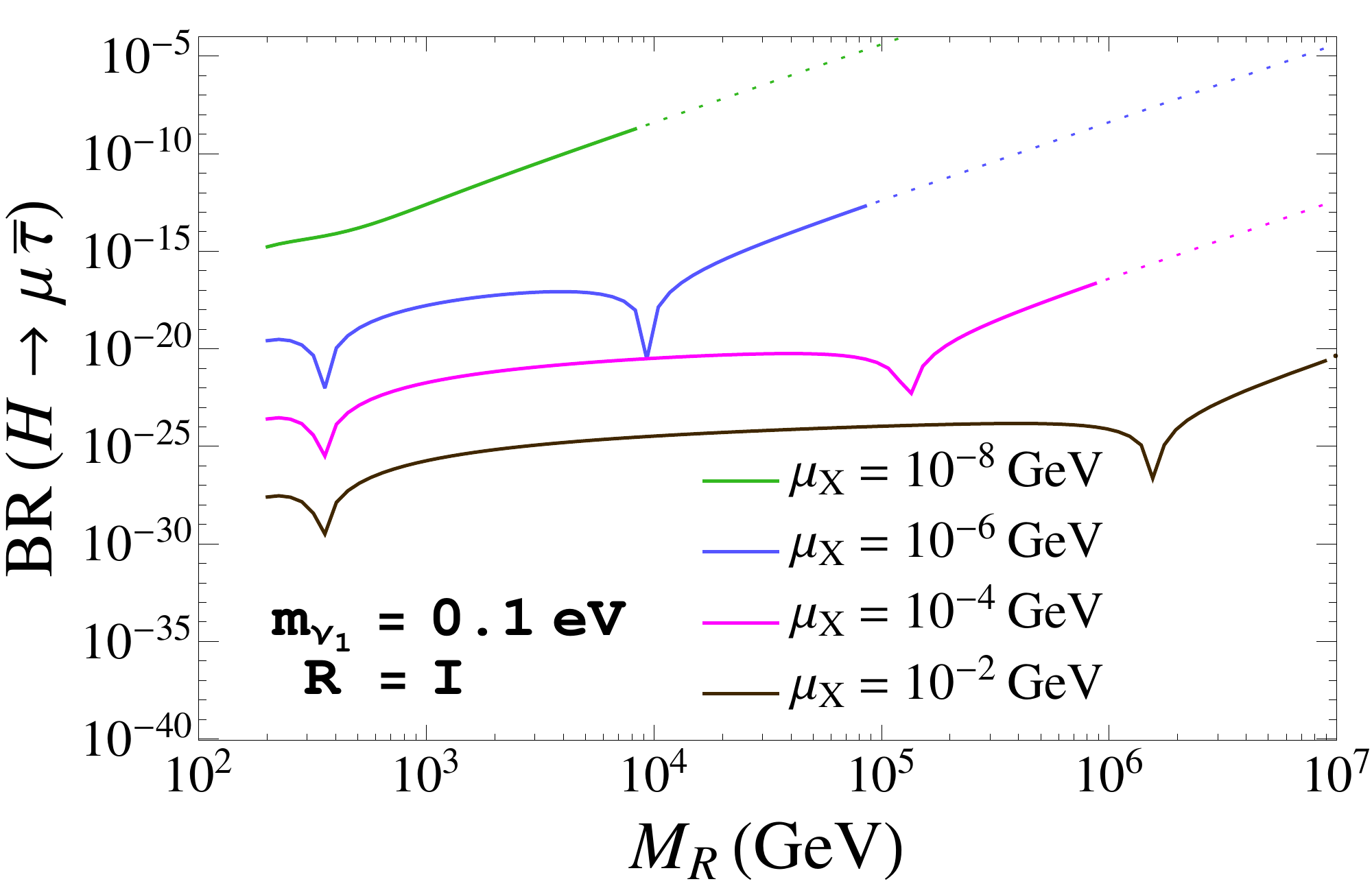} &
\includegraphics[width=0.49\textwidth]{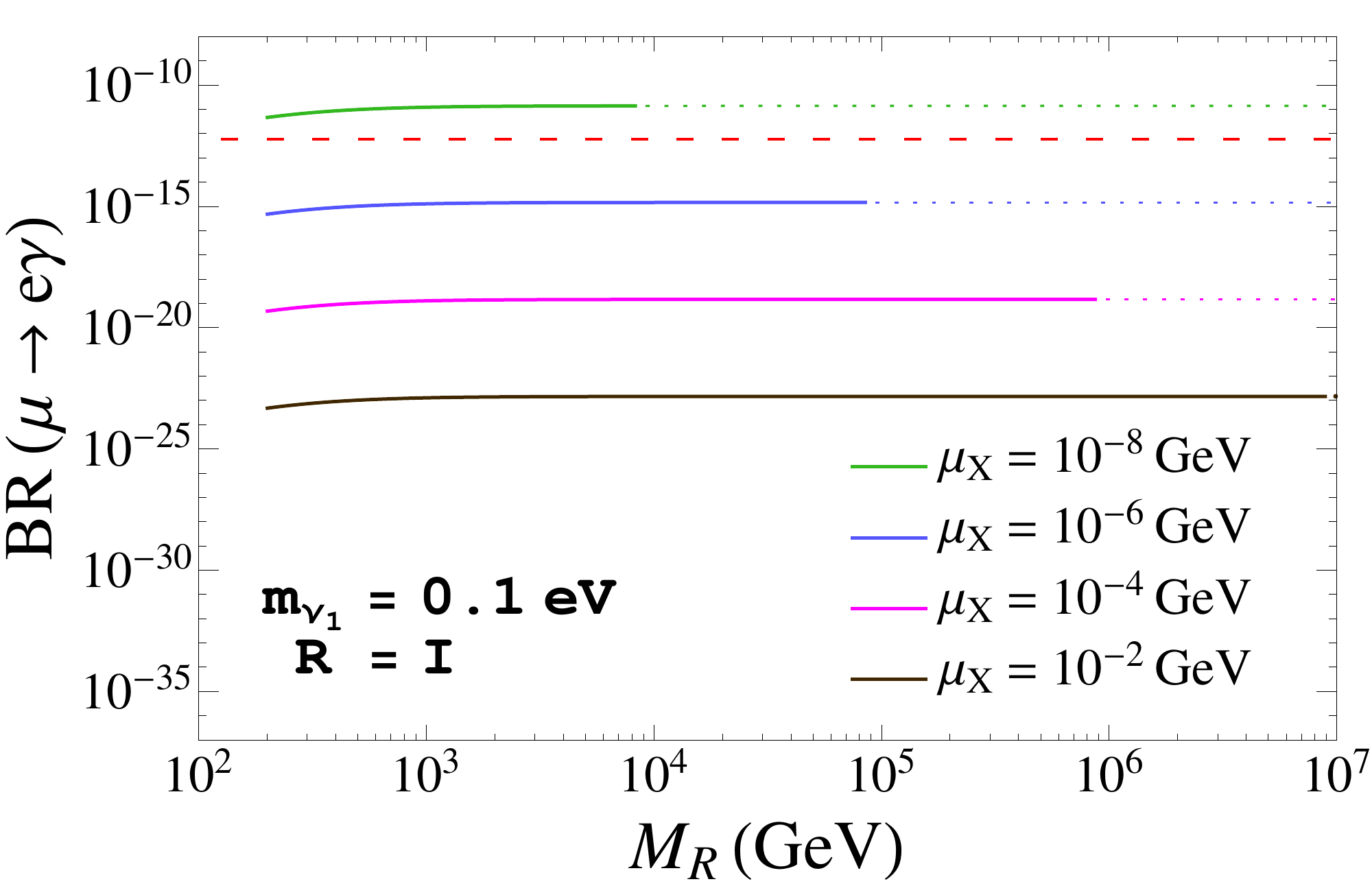}
\end{tabular}
\caption{Branching ratios of $H \to \mu\overline\tau$ (left panel) and $\mu \to e \gamma$ (right panel) as functions of $M_R$ for different values
of $\mu_X=(10^{-8},10^{-6},10^{-4},10^{-2})$ GeV from top to bottom. In both panels, $m_{\nu_1} = 0.1$ eV and $R = I$. The horizontal red dashed line
denotes the current experimental upper bound for $\mu \to e \gamma$, BR($\mu \to e \gamma$) $< 5.7 \times 10^{-13}$~\cite{Adam:2013mnn}. Dotted lines represent
nonperturbative neutrino Yukawa couplings.}\label{BRs_MR_degenerate}
\end{center}
\end{figure}

Next we study the sensitivity in the LFV rates to other choices of $\mu_X$. For this study we focus on the largest LFVHD rates, BR($H \to \mu\bar\tau$), and on the most
constraining BR$(\mu \to e \gamma)$ rates. In
fig.~\ref{BRs_MR_degenerate} we show the predictions for the LFV rates for different values of $\mu_X=(10^{-8},10^{-6},10^{-4},10^{-2})$ GeV. The other input parameters have been
fixed here to $m_{\nu_1} = 0.1$ eV and $R = I$. 
On the left panel of fig.~\ref{BRs_MR_degenerate} we see again the increase of BR($H \to \mu\overline\tau$)  as $M_R$ grows, which is more pronounced in the region where $M_R$
is large and $\mu_X$ is low, and, therefore, where the Yukawa couplings are large [see eq.~(\ref{CasasIbarraISS})]. We have checked that, in that region, the dominant diagrams are
by far the divergent diagrams (1), (8), and (10), and that the BR($H \to \mu\overline\tau$) rates  grow as $M_R^4$.
In this plot, as well as in the previous plot of BR($H\to l_k\bar l_m$) in fig.~\ref{ALL_LFVdecays}, we can also identify the appearance of different dips, which we have
understood as destructive interferences among the various contributing diagrams. More precisely, we have checked that the dips that appear at large $M_R$ and just before
the $M_R^4$ growing region are due to partial cancellations between diagrams (1),(8), and (10), while the other dips that appear at lower $M_R$ happen among
diagrams (2)-(6) [(7) and (9) are subleading]. 
These last diagrams have relevant contributions to BR($H \to \mu\overline\tau$)  only for low values of the Yukawa couplings.
We also observe a fast growth of the LFV Higgs rates as $\mu_X$ decreases from $10^{-2}$ GeV to $10^{-8}$ GeV. However, not all the values of $M_R$ and $\mu_X$ are allowed,
because they may generate nonperturbative Yuwaka entries, expressed again in this figure by dotted lines. Therefore, the largest LFV Higgs rates permitted by our perturbativity
requirements [eq.~(\ref{Ymax})] are approximately BR($H \to \mu\overline\tau$) $\sim 10^{-9}$, obtained for $\mu_X = 10^{-8}$ GeV and $M_R \simeq 10^4$ GeV. Larger
values of $M_R$, for this choice of $\mu_X$, would produce Yukawa couplings that are not perturbative.

Nevertheless, we must pay attention to the predictions of BR($\mu \to e \gamma$) for this choice of parameters, because they can be excluded by its quite restrictive present
experimental upper bound, as shown on the right panel of fig.~\ref{BRs_MR_degenerate}. In this plot, the dependence of BR($\mu \to e \gamma$) on $M_R$ is depicted, for the
same choices of $\mu_X$, $m_{\nu_1}$, and $R$ as in the left panel. The horizontal red dashed line denotes again its current upper bound; see eq.~(\ref{MUEGmax}). In addition to
what we have already learned about the approximate behavior of the BR($\mu \to e \gamma$) rates going as $|(Y_\nu Y_\nu^\dag)_{12}/M_R^2|^2$, which explains the constant behavior
with $M_R$, we also learn from this figure about the generic behavior with $\mu_X$,
which leads to increasing LFV rates for decreasing $\mu_X$ values, for both LFVHD and radiative processes. In particular, we see that
small values of $\mu_X \leq \mathcal O(10^{-8}~\rm{GeV})$ lead to BR($\mu \to e \gamma$) rates that are excluded by the present experimental upper bound.
Taking this into account, the largest value of BR($H \to \mu\overline\tau$), for the choice of parameters fixed in fig.~\ref{BRs_MR_degenerate}, that is allowed by
the BR($\mu \to e \gamma$) upper bound (this being more restrictive than the perturbativity requirement in this case) is $\sim 10^{-12}$, which is obtained for $M_R = 10^5$ GeV
and $\mu_X = 10^{-6}$ GeV.

\begin{figure}[t!]
\begin{center}
\begin{tabular}{cc}
\includegraphics[width=0.49\textwidth]{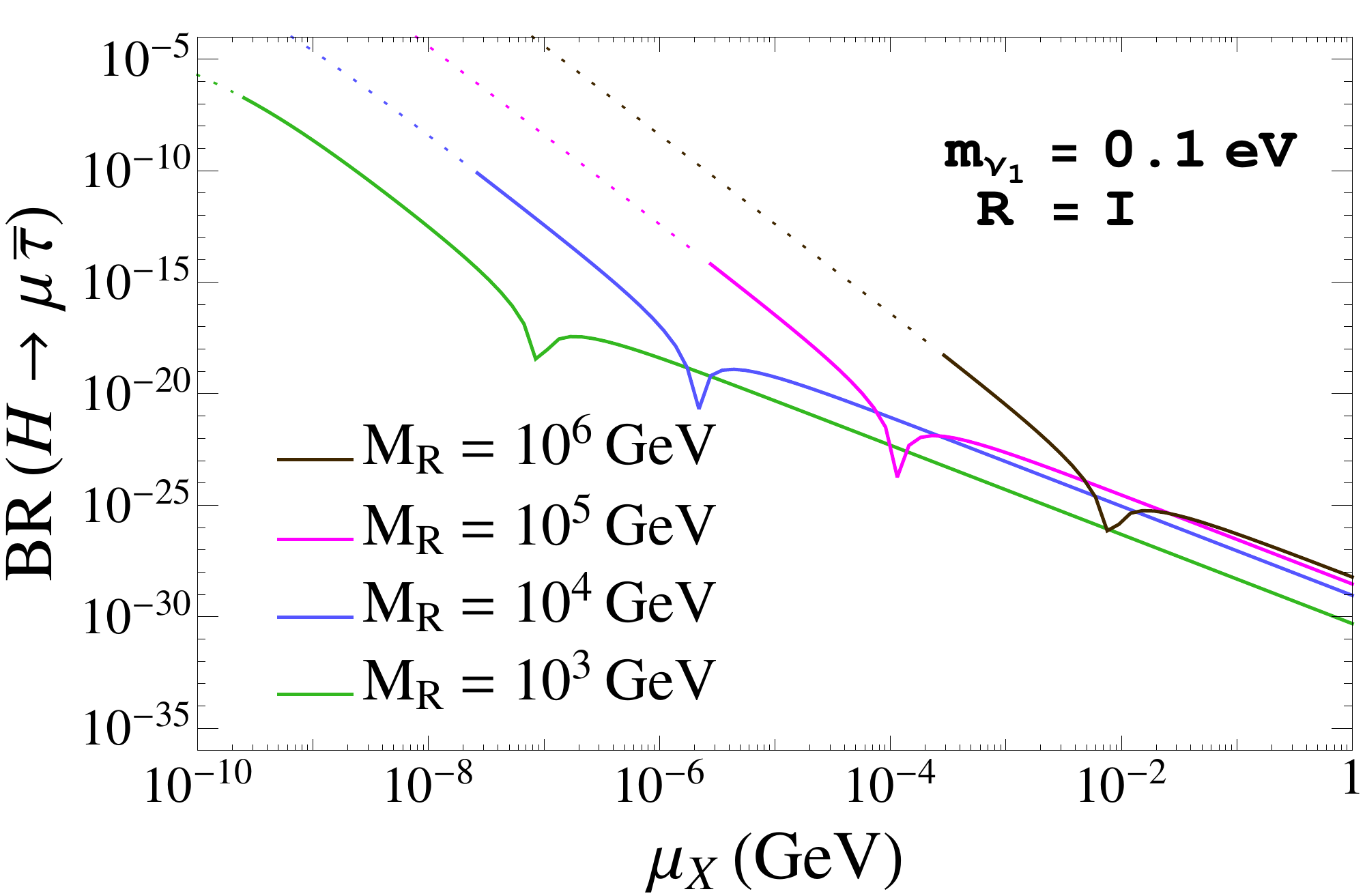} &
\includegraphics[width=0.49\textwidth]{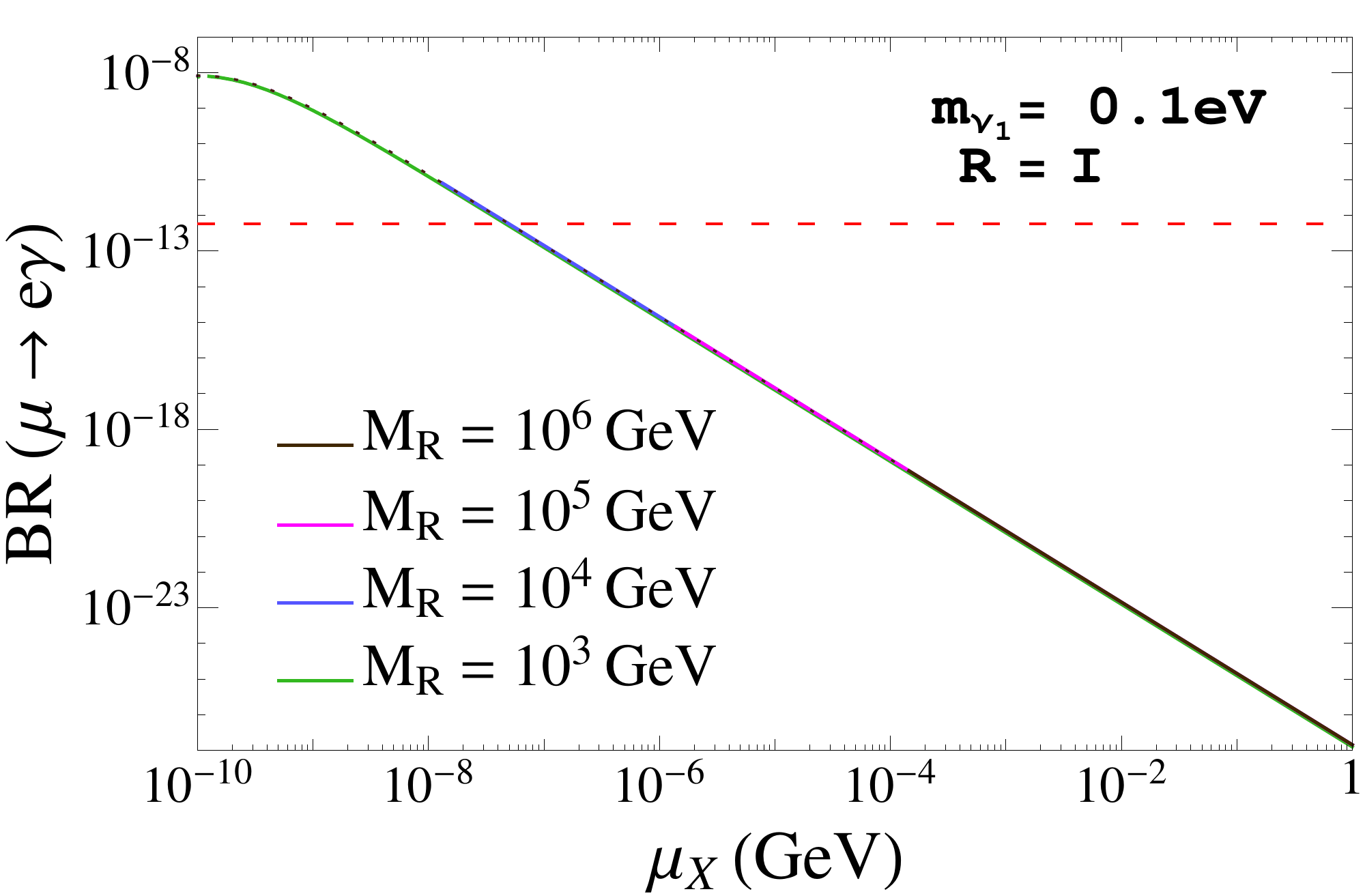}
\end{tabular}
\caption{Branching ratios of $H \to \mu\overline\tau$ (left panel) and $\mu \to e \gamma$ (right panel) as functions of $\mu_X$ for
different values of $M_R=(10^{6},10^{5},10^{4},10^{3})$ GeV from top to bottom. In both panels, $m_{\nu_1} = 0.1$ eV and $R = I$. The horizontal
red dashed line denotes the current experimental upper bound for $\mu \to e \gamma$, BR($\mu \to e \gamma$) $< 5.7 \times 10^{-13}$~\cite{Adam:2013mnn}.
Dotted lines represent nonperturbative neutrino Yukawa couplings.}\label{BRs_muX_degenerate}
\end{center}
\end{figure}

The behavior of BR($H \to \mu\overline\tau$) and BR($\mu \to e \gamma$) as functions of $\mu_X$, for several values of $M_R$, $m_{\nu_1} = 0.1$ eV, and $R = I$, is displayed
in fig.~\ref{BRs_muX_degenerate}. As already seen in fig.~\ref{BRs_MR_degenerate}, both LFV rates decrease as $\mu_X$ grows; however, the functional dependence is not the same.
The LFV radiative decay rates decrease as $\mu_X^{-2}$, in agreement with the approximate expression (\ref{approxformula}), while the LFVHD rates go as $\mu_X^{-4}$ when the Yukawa
couplings are large.
For a fixed value of $\mu_X$, the larger $M_R$ is, the larger BR($H \to \mu\overline\tau$) can be, while the prediction for BR($\mu \to e \gamma$) is the same
for any value of $M_R$. We have already learned this independence of the LFV radiative decays on $M_R$ from the previous figure, which can be easily confirmed on the right panel
of fig.~\ref{BRs_muX_degenerate}, where all the lines for different values of $M_R$ are superimposed. We observe again the existence of dips in the left panel
of fig.~\ref{BRs_muX_degenerate}. We also see in this figure that
the smallest value of $\mu_X$ allowed by the BR($\mu \to e \gamma$) upper bound is $\mu_X\sim 5\times10^{-8}$ GeV, which is directly translated to
a maximum allowed value of BR($H \to \mu\overline\tau$) $\sim 10^{-11}$, for $M_R = 10^4$ GeV.

\begin{figure}[t!]
\begin{center}
\begin{tabular}{cc}
\includegraphics[width=0.49\textwidth]{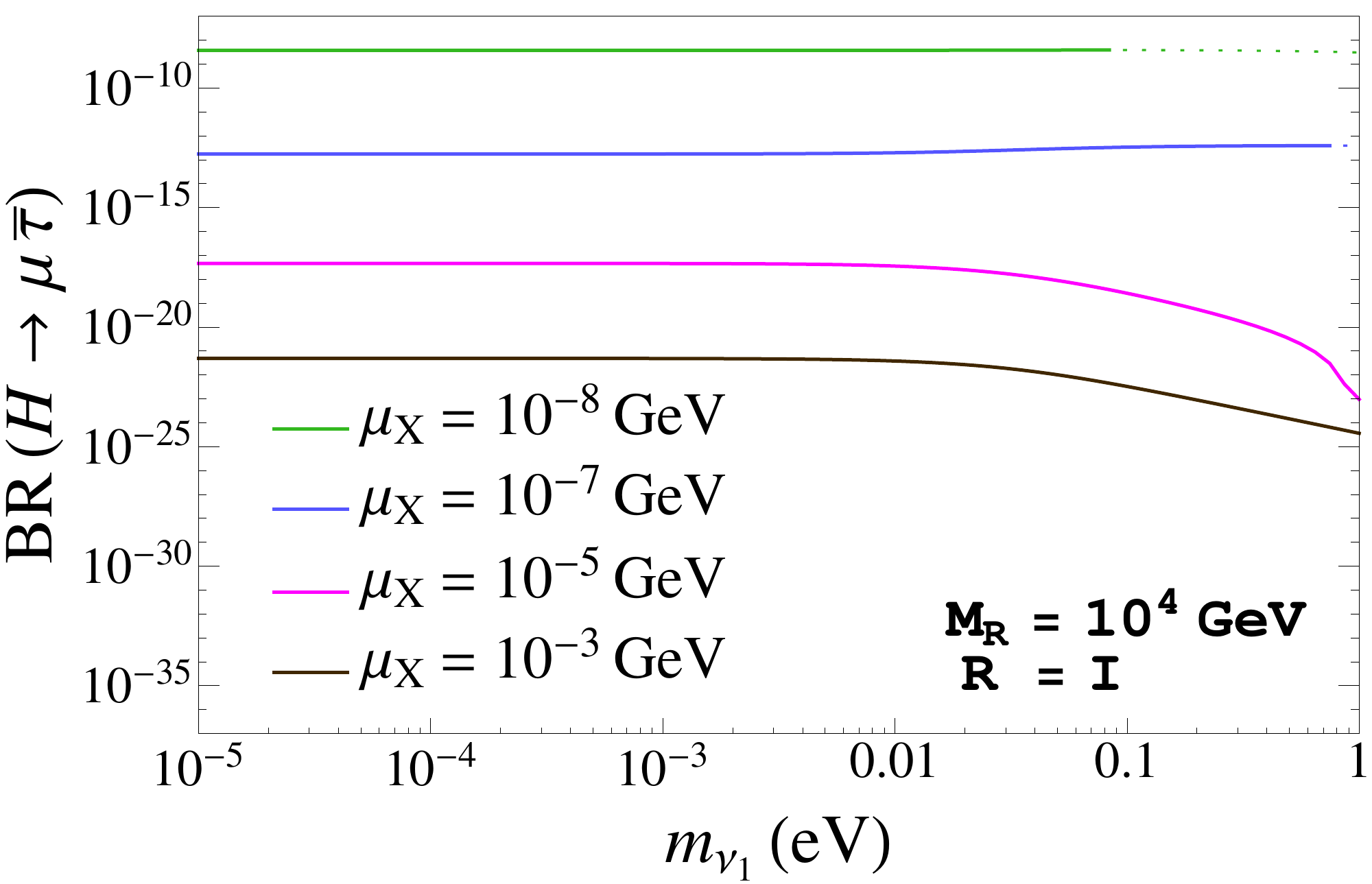} &
\includegraphics[width=0.49\textwidth]{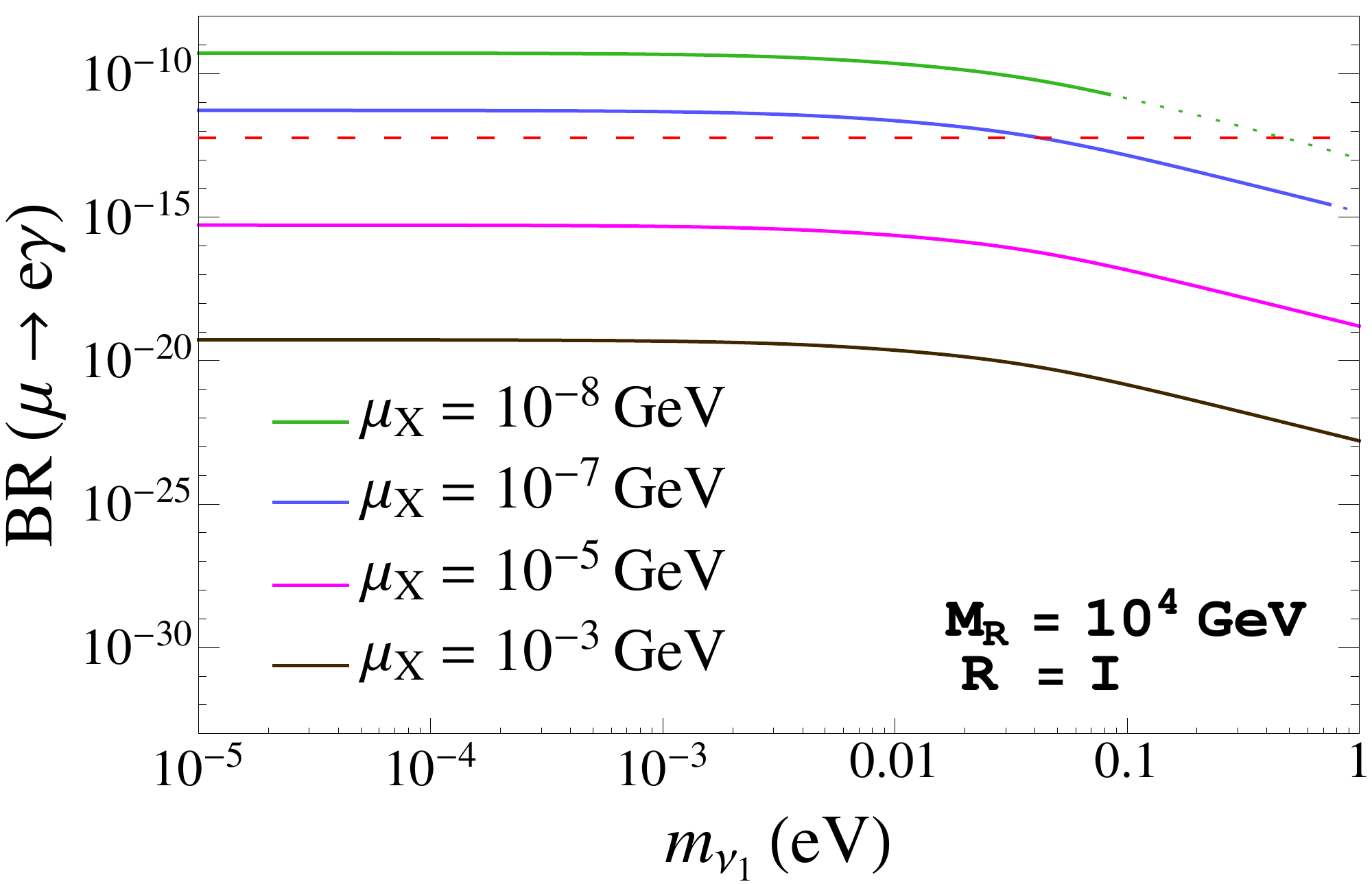}
\end{tabular}
\caption{Branching ratios of $H \to \mu\overline\tau$ (left panel) and $\mu \to e \gamma$ (right panel) as functions of $m_{\nu_1}$ for
different values of $\mu_X=(10^{-8},10^{-7},10^{-5},10^{-3})$ GeV from top to bottom. In both panels, $M_R = 10^4$ GeV and $R = I$. The horizontal red dashed
line denotes the current experimental upper bound for $\mu \to e \gamma$, BR($\mu \to e \gamma$) $< 5.7 \times 10^{-13}$~\cite{Adam:2013mnn}. Dotted lines represent
nonperturbative neutrino Yukawa couplings [see eq.~(\ref{Ymax})].}\label{BRs_mnu1_degenerate}
\end{center}
\end{figure}

The dependence of BR($H \to \mu\overline\tau$) and BR($\mu \to e \gamma$) on the lightest neutrino mass $m_{\nu_1}$ is studied in fig.~\ref{BRs_mnu1_degenerate}, for several
values of $\mu_X$ with $M_R = 10^4$ GeV and $R = I$. For the chosen parameters in this figure, a similar dependence on $m_{\nu_1}$ is observed in both observables, in which there
is a flat behavior with $m_{\nu_1}$ except for values of $m_{\nu_1} \gtrsim$ 0.01 eV. For these values, the LFV rates decrease as $m_{\nu_1}$ grows. 

The behavior of BR($l_m\to l_k\gamma$) with $m_{\nu_1}$ can be understood from the fact that the flavor violation arises from the nondiagonal  terms of $(Y_\nu Y_\nu^\dag)_{km}$. In
the simplified case of real $R$ and $U_{\rm PMNS}$ matrices, for diagonal and degenerate $M_R$ and $\mu_X$, and by using eqs.~(\ref{heavymasses}) and~(\ref{CasasIbarraISS}), we
find the following simple expression for the nondiagonal $km$ elements:

\begin{equation}
\frac{v^2\Big(Y_\nu Y_\nu^\dag\Big)_{km}}{M_R^2}\approx
\left\{\begin{array}{ll}
\displaystyle{\frac{1}{\mu_X}\Big(U_{\rm PMNS} \sqrt{\Delta m^2}~ U_{\rm PMNS}^T\Big)_{km}} & ,\,{\rm for}~m_{\nu_1}^2\ll|\Delta m^2_{ij}| \,,\\\\
\displaystyle{\frac{1}{\mu_X}\frac{\Big(U_{\rm PMNS} \Delta m^2~ U_{\rm PMNS}^T\Big)_{km}}{2m_{\nu_1}} }& ,\,{\rm for}~ m_{\nu_1}^2\gg|\Delta m^2_{ij}|\,,
\end{array}\right.
\label{YnuYnudependence}
\end{equation}
where we have defined
\begin{equation}\label{deltam}
\Delta m^2\equiv \rm{diag}(0, \Delta m_{21}^2,\Delta m_{31}^2)~,
\end{equation}
and we have expanded properly $m_{\nu_2}$ and $m_{\nu_3}$ in eq.~(\ref{mnu2mnu3}) in terms of $m_{\nu_1}$ and $\Delta m_{ij}^2$.
Therefore, using eqs.~(\ref{approxformula})-(\ref{deltam}), we conclude that the BR($\mu\to~e\gamma$) rates have a flat behavior
with $m_{\nu_1}$ for low values of $m_{\nu_1}\lesssim 0.01~\rm{eV}$, but they decrease with $m_{\nu_1}$ for larger values, explaining the observed behavior
in fig.~\ref{BRs_mnu1_degenerate}.

\begin{figure}[t!] 
\begin{center}
\includegraphics[width=.6\textwidth]{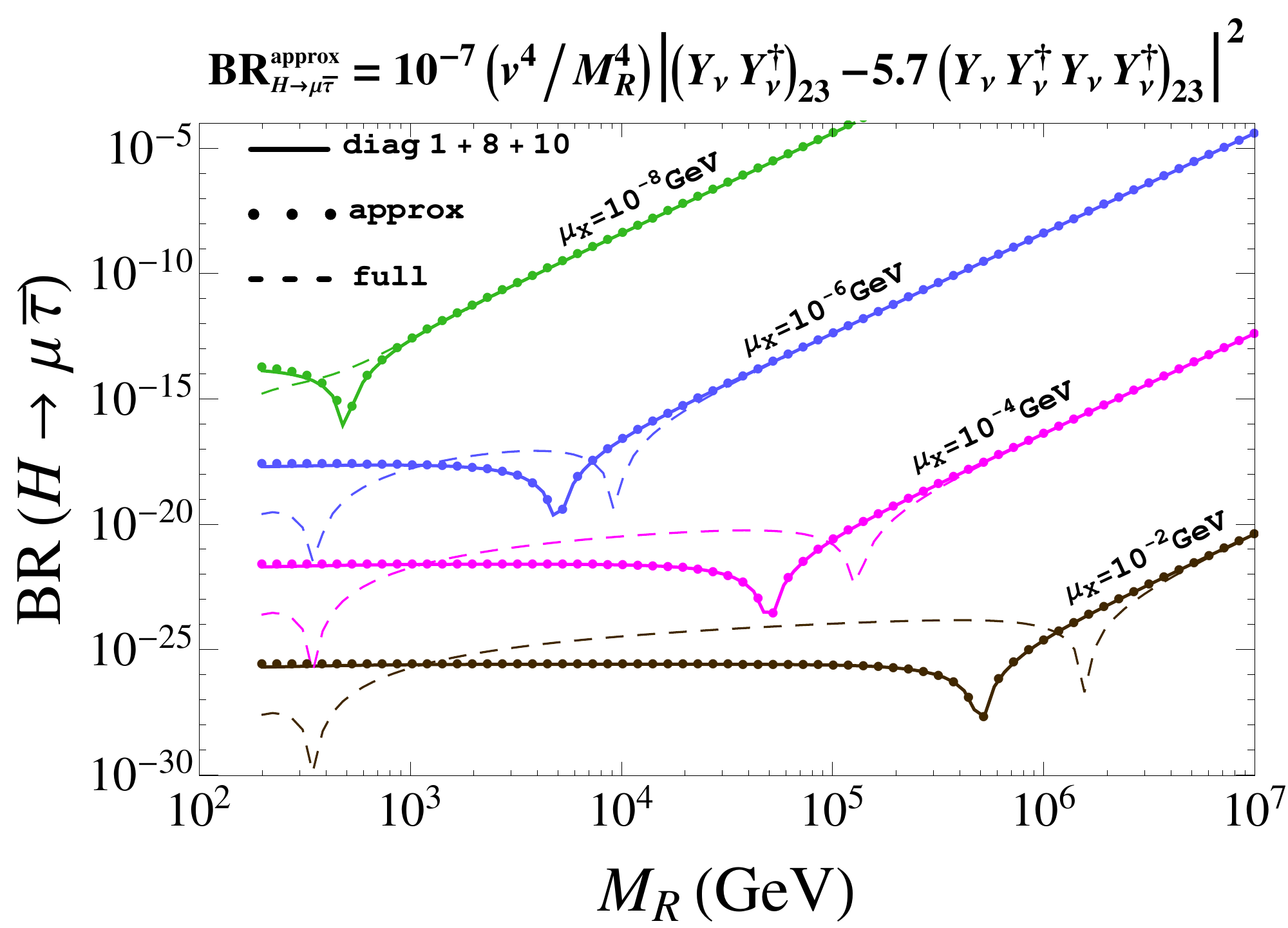} 
\caption{Comparison between the predicted rates for BR($H\to\mu\overline{\tau}$) taking (1) the full one-loop formulas (dashed lines); (2) just
the contributions from diagrams (1), (8), and (10) of fig.~\ref{diagrams} (solid lines); and (3) the approximate formula of eq.~(\ref{FIThtaumu}) (dotted lines)}
\label{fitLFVHD}
\end{center}
\end{figure}

By taking into account all the behaviors learned above, we have tried to find an approximate simple formula that could explain the main features of the
BR($H \to \mu\overline\tau$) rates. As we have already said, in contrast to what we have seen for the LFV radiative decays in eq.~(\ref{approxformula}), a simple functional
dependence being proportional to $|(Y_\nu Y_\nu^\dagger)_{23}|^2$ is not enough to describe our results for the BR($H \to \mu\overline\tau$) rates.
Considering that, in the region where the Yukawa couplings are large, the LFVHD rates are dominated by diagrams (1), (8), and (10), we have looked for a simple expression
that could properly fit the contributions from these dominant diagrams. From this fit we have found the following approximate formula:
\begin{equation}\label{FIThtaumu}
{\rm BR}^{\rm approx}_{H\to\mu\bar\tau}=10^{-7}\frac{v^4}{M_R^4}~\Big|(Y_\nu Y_\nu^\dagger)_{23}-5.7(Y_\nu Y_\nu^\dagger Y_\nu Y_\nu^\dagger)_{23}\Big|^2,
\end{equation}
which turns out to work reasonably well. In fig.~\ref{fitLFVHD} we show   
the predicted rates of BR($H\to\mu\overline{\tau}$) with (1) the full one-loop formulas (dashed lines); (2) taking just the contributions from
diagrams (1), (8), and (10) of fig.~\ref{diagrams} (solid lines); and (3) using eq.~(\ref{FIThtaumu}) (dotted lines). We see clearly that this eq.~(\ref{FIThtaumu}) reproduces
extremely well the contributions from  diagrams (1), (8), and (10) and approximates reasonably well the full rates. The approximation is pretty good indeed for the $M_R$ region
above the dips. The change of functional behavior with $M_R$ in the two different $M_R$ regions, from nearly flat with $M_R$ in the approximate result to fast growing as $\sim M_R^4$,
also gives a reasonable approach to the full result, as well as the appearance of dips. The location of the dips is however not so accurately described by the approximate formula,
since in the region where the cancellation among the diagrams (1), (8), and (10) takes place the other diagrams (not considered in the fit) also contribute. Overall, we 
find the approximate formula  given by eq.~(\ref{FIThtaumu}) very useful for generic estimates in the ISS, which could also be applied to other parametrizations of the neutrino
Yukawa couplings.      

This particular choice for the fitting function can be easily understood using the electroweak interaction basis of eq.~(\ref{LagrangianISS}) and applying the mass insertion
approximation (MIA). Looking at the finite contribution coming from diagrams (1), (8), and (10), we can see that, at the lowest order in the MIA, the Higgs decay amplitude has
a similar behavior to the dimension-six operator that governs the LFV radiative decays, which is proportional to
\begin{equation}\label{radiative}
 \frac{v^2(Y_\nu Y_\nu^\dagger)_{km}}{M_R^2}\,.
\end{equation}
However, there are other contributions to the Higgs decays that are not present in the case of the radiative decays, owing to the different chiral structure of the lepton flavor
violating operators.
For example, having two mass insertions of $LR$ type, one in each internal neutrino line of a loop like that of diagram~1, will give a contribution to the amplitude proportional to
\begin{equation}\label{Ynu4}
 \frac{v^2(Y_\nu Y_\nu^\dagger Y_\nu Y_\nu^\dagger)_{km}}{M_R^2}\,.
\end{equation}
Then, using again eqs.~(\ref{CasasIbarraISS}) and~(\ref{heavymasses}), we find the following simple expression:
\begin{equation}\label{Ynu4dependence}
 \frac{v^2(Y_\nu Y_\nu^\dagger Y_\nu Y_\nu^\dagger)_{km}}{M_R^2}=\frac{M_R^2}{v^2\mu_X^2}\Big(U_{\rm PMNS} \Delta m^2~ U_{\rm PMNS}^T\Big)_{km}\,.
\end{equation}
Thus, we can clearly see from the above result that the second contribution in eq.~(\ref{FIThtaumu}) is the one that dominates at large $M_R$ and low $\mu_X$, i.e., at large
Yukawa couplings, and, indeed, it reproduces properly the behavior of BR($H\to \mu\overline\tau$) in this limit, with BR $\propto M_R^4/\mu_X^{4}$. It is also independent
of $m_{\nu_1}$, explaining the flat behavior in fig.~\ref{BRs_mnu1_degenerate} for low values of $\mu_X$.
Moreover, if the two contributions in eq.~(\ref{FIThtaumu}) have opposite signs, they will interfere destructively, leading to a dip in the decay rate when both contributions
are of the same size.
From eqs.~(\ref{YnuYnudependence}) and~(\ref{Ynu4dependence}), we can deduce that the position of the dip should verify $M_R^{-2} \mu_X  \sim \mathrm{constant}$, which is
the behavior observed at large $M_R$ in figs.~\ref{BRs_MR_degenerate}-\ref{BRs_muX_degenerate}. The other dips, which appear for $M_R \simeq 300\,\mathrm{GeV}$ in
fig.~\ref{BRs_MR_degenerate}, come from a destructive interference between the other diagrams, as we have said. 

\begin{figure}[t!]
\begin{center}
\begin{tabular}{cc}
\includegraphics[width=.49\textwidth]{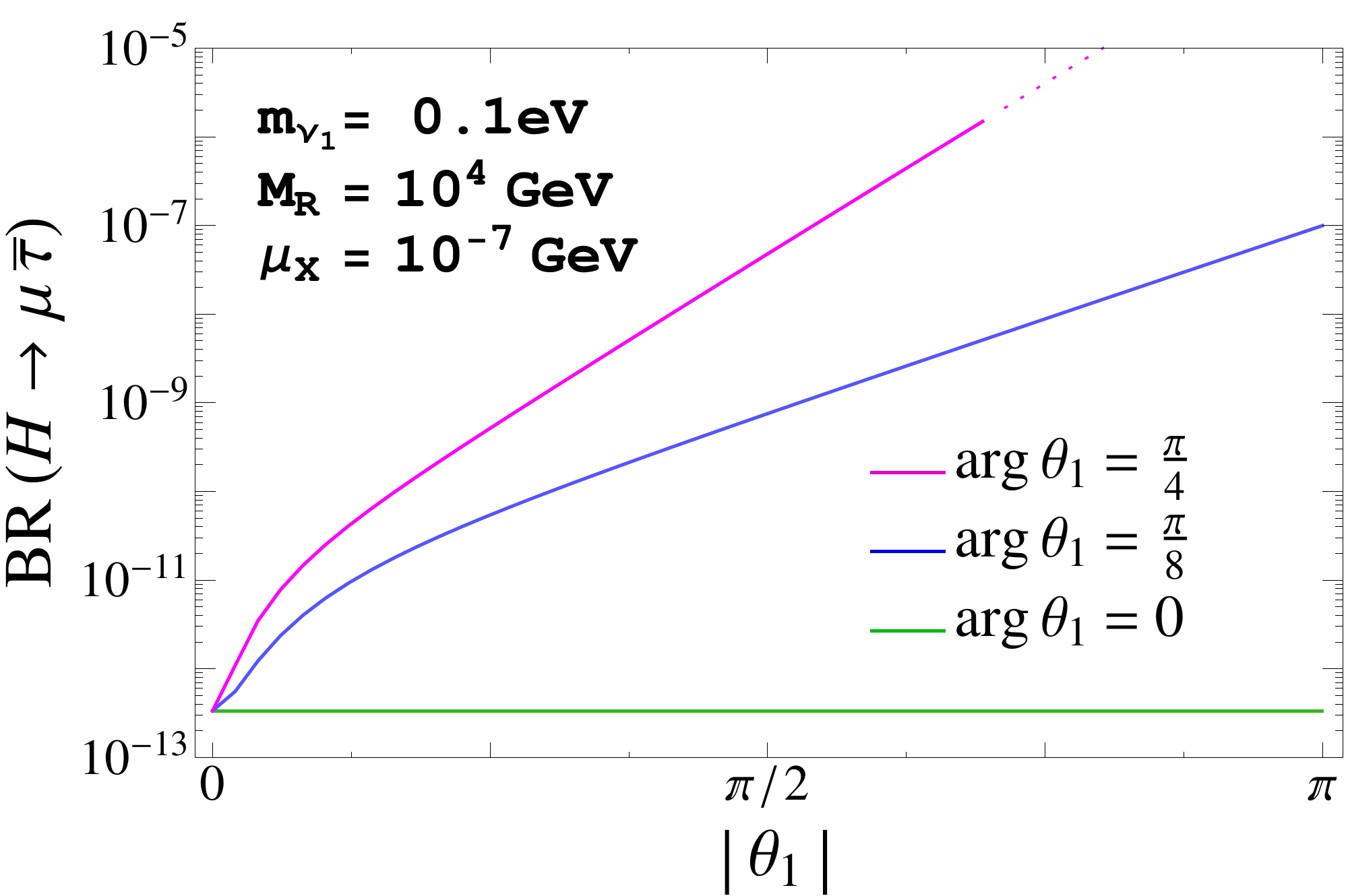} &
\includegraphics[width=.49\textwidth]{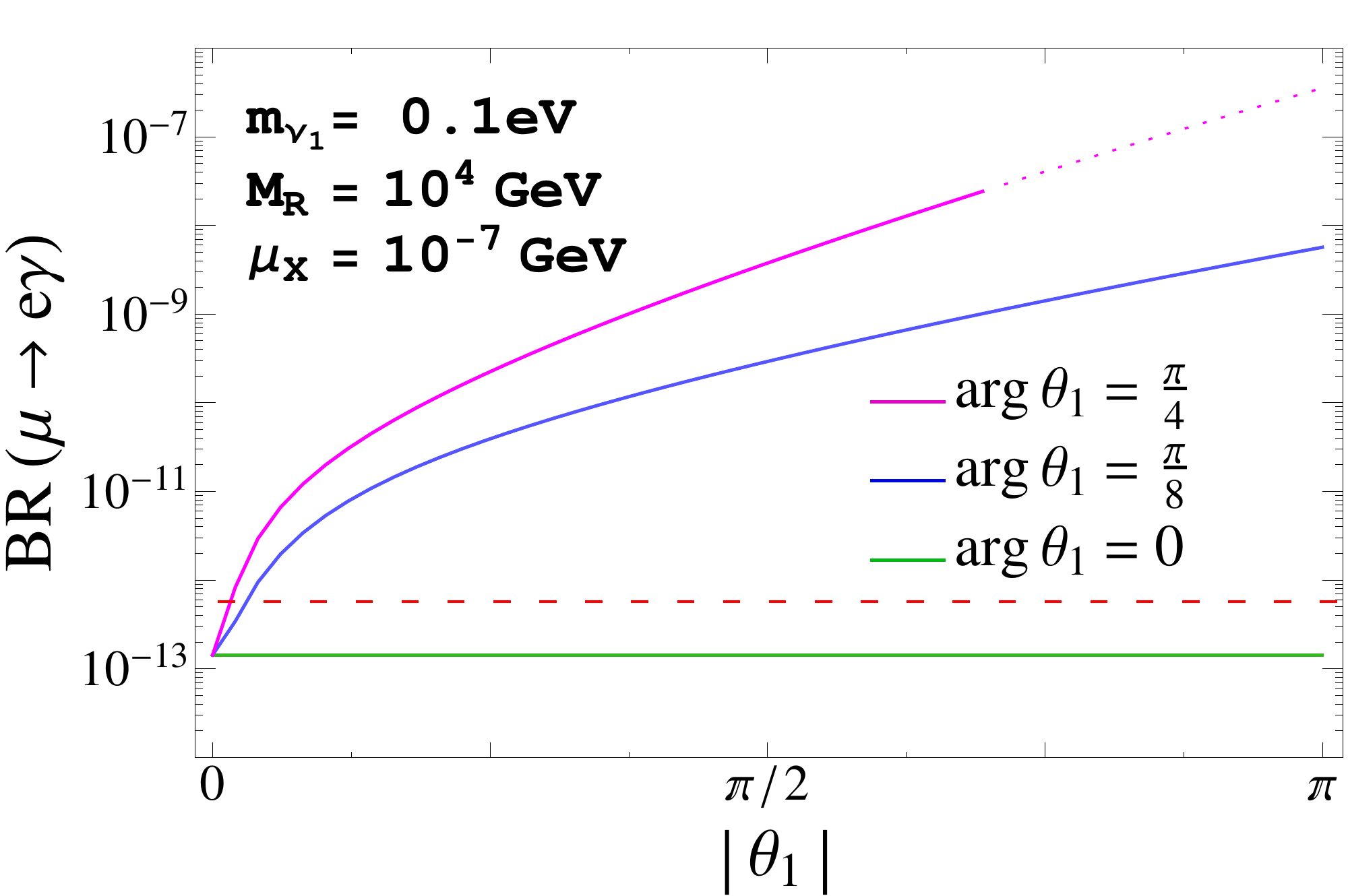}
\end{tabular}
\caption{Branching ratios of $H \to \mu\overline\tau $ (left panel) and $\mu \to e \gamma$ (right panel) as functions of $|\theta_1|$ for different values
of arg$\theta_1$. In both panels, $M_R = 10^4$ GeV, $\mu_X = 10^{-7}$ GeV, and $m_{\nu_1} =$ 0.1 eV. The horizontal red dashed line denotes the current experimental
upper bound for $\mu \to e \gamma$, BR($\mu \to e \gamma$) $< 5.7 \times 10^{-13}$~\cite{Adam:2013mnn}. Dotted lines represent nonperturbative neutrino Yukawa
couplings [see eq.~(\ref{Ymax})].}\label{BRs_theta1_degenerate}
\end{center}
\end{figure}
Next, we display in fig.~\ref{BRs_theta1_degenerate} the dependence of the $H \to \mu\overline\tau$ and $\mu \to e \gamma$ decay rates on $|\theta_1|$ for different
values of arg$\theta_1=0,\pi/8, \pi/4$, with $M_R = 10^4$ GeV, $\mu_X = 10^{-7}$ GeV, and $m_{\nu_1} =$ 0.1 eV. 
First of all, we highlight the flat behavior of both LFV rates with $|\theta_1|$ for real $R$ matrix (arg$\theta_1$ = 0), which is a direct consequence of the 
degeneracy of $M_R$ and $\mu_X$. In other words, the LFV rates for the degenerate heavy neutrinos case are independent of $R$ if it is real. Once we abandon the real case 
and consider values of arg$\theta_1$ different from zero, a strong dependence on $|\theta_1|$ appears. The larger $|\theta_1|$ and/or arg$\theta_1$ are, the larger the LFV 
rates become. 
On the other hand, only values of $|\theta_1|$ lower than $\pi/32$ with arg$\theta_1 = \pi/8$ in this figure are allowed by the $\mu \to e \gamma$ constraint, which allows us
to reach values of BR($H \to \mu \bar \tau$) $\sim 10^{-12}$ at the most. We have also explored the LFV rates as functions of complex $\theta_2$ and $\theta_3$ and we have
reached similar conclusions as for $\theta_1$. Therefore, by choosing complex $\theta_{1,2,3}$ the LFV Higgs decay rates that are allowed by the upper bounds on the radiative
decays do not increase with respect to the real case, which is equal to the previous $R=I$ reference case, due to the independence on real $R$, as we have already said.

\begin{figure}[t!] 
\begin{center}
\begin{tabular}{cc}
\hspace{-0.5cm}\includegraphics[width=.6\textwidth]{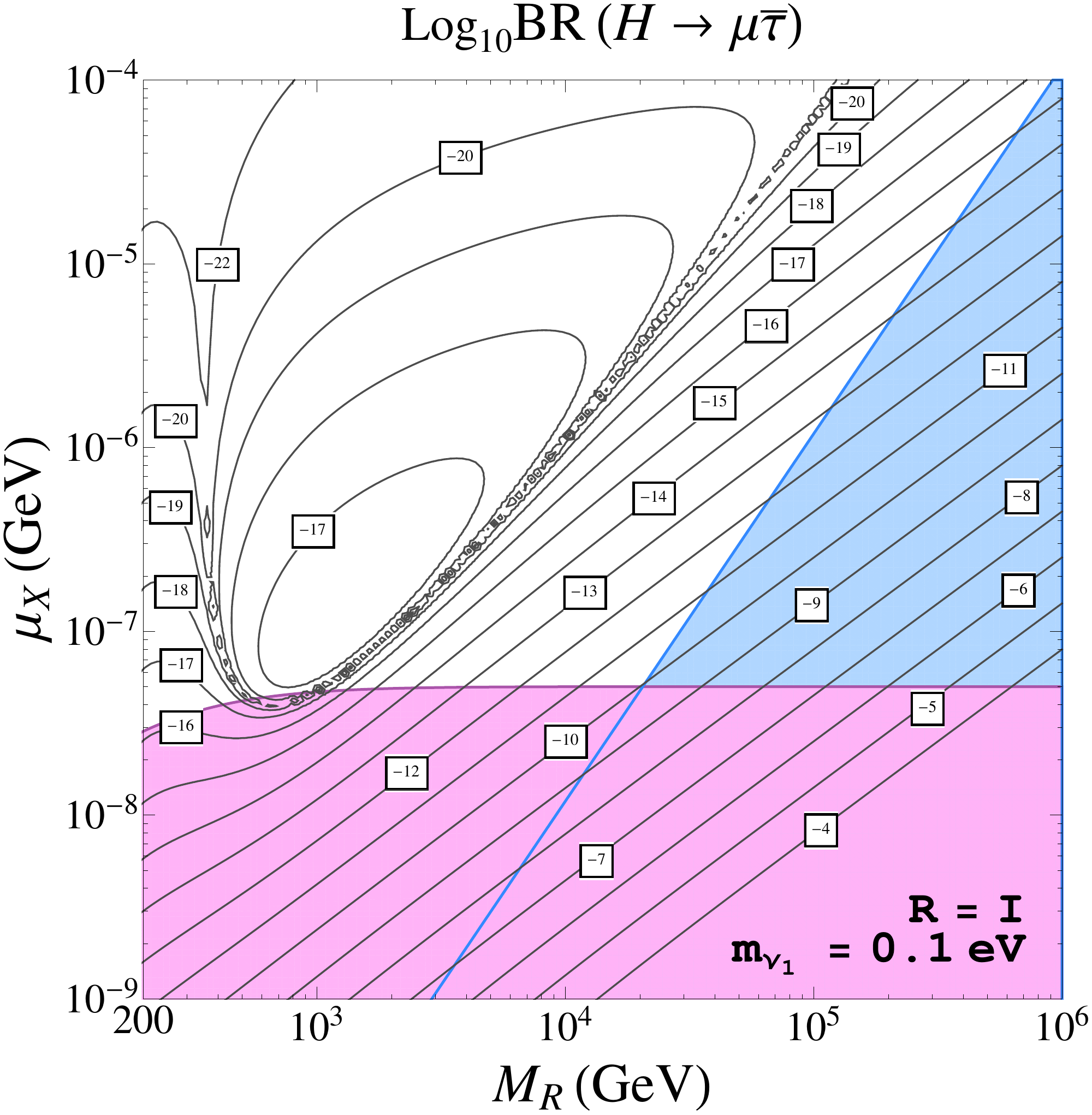} &
\end{tabular}
\caption{Contour lines of BR($H \to \mu \bar \tau$) in the $(M_R,\mu_X)$ plane for $R=I$ and $m_{\nu_1}=0.1\,{\rm eV}$. The horizontal area in pink is excluded by the upper bound
on  BR$(\mu \to e \gamma)$. The oblique area in blue is excluded by the perturbativity requirement of the neutrino Yukawa couplings.}
\label{ContourPlot_degenerate}
\end{center}
\end{figure}

Once we have studied the behavior of all the LFV observables considered here with the most relevant parameters, we next present the results for the maximum allowed LFV Higgs
decay rates in the case of heavy degenerate neutrinos. The  plot in fig.~\ref{ContourPlot_degenerate} shows the contour lines of BR($H \to \mu \bar \tau$) in
the $(M_R,\mu_X)$ plane for $R=I$ and $m_{\nu_1}=0.1\,{\rm eV}$. The horizontal area in pink is excluded by not respecting the present upper bound on  BR$(\mu \to e \gamma)$.
The oblique area in blue is excluded by not respecting  the perturbativity of the neutrino Yukawa couplings. These contour lines summarize the previously learned behavior
with $M_R$ and $\mu_X$, which lead to the largest values for the LFVHD rates in the bottom right-hand corner of the plot, i.e., at large $M_R$ and small $\mu_X$. We also notice
the appearance of dips in the $(M_R,\mu_X)$ plane that correspond to the previously commented dips in the previous figures. The most important 
conclusion from this contour plot
is that the maximum allowed LFVHD rate is approximately BR$(H \to \mu \bar \tau)\sim 10^{-10}$ and it is found for $M_R\sim 2\times 10^4\,{\rm GeV}$ and
$\mu_X \sim  5\times 10^{-8} \, {\rm GeV}$. We have found similar conclusions for BR$(H \to e \bar \tau)$.  

\subsection{Hierarchical heavy neutrinos}
\label{hierarchical}
The case of hierarchical heavy neutrinos refers here to hierarchical masses among generations and it is implemented  by choosing hierarchical entries in the
$M_R={\rm diag}(M_{R_1},M_{R_2},M_{R_3})$ matrix. As for the $\mu_X={\rm diag}(\mu_{X_1},
\mu_{X_2},\mu_{X_3})$ matrix that introduces the tiny splitting within the heavy masses in the same generation we choose it here to be degenerate, $\mu_{X_{1,2,3}}=\mu_X$.
We focus on the normal hierarchy $M_{R_1}<M_{R_2}<M_{R_3}$, since we have found similar conclusions for other hierarchies. 

\begin{figure}[t!]
\begin{center}
\begin{tabular}{cc}
\includegraphics[width=0.49\textwidth]{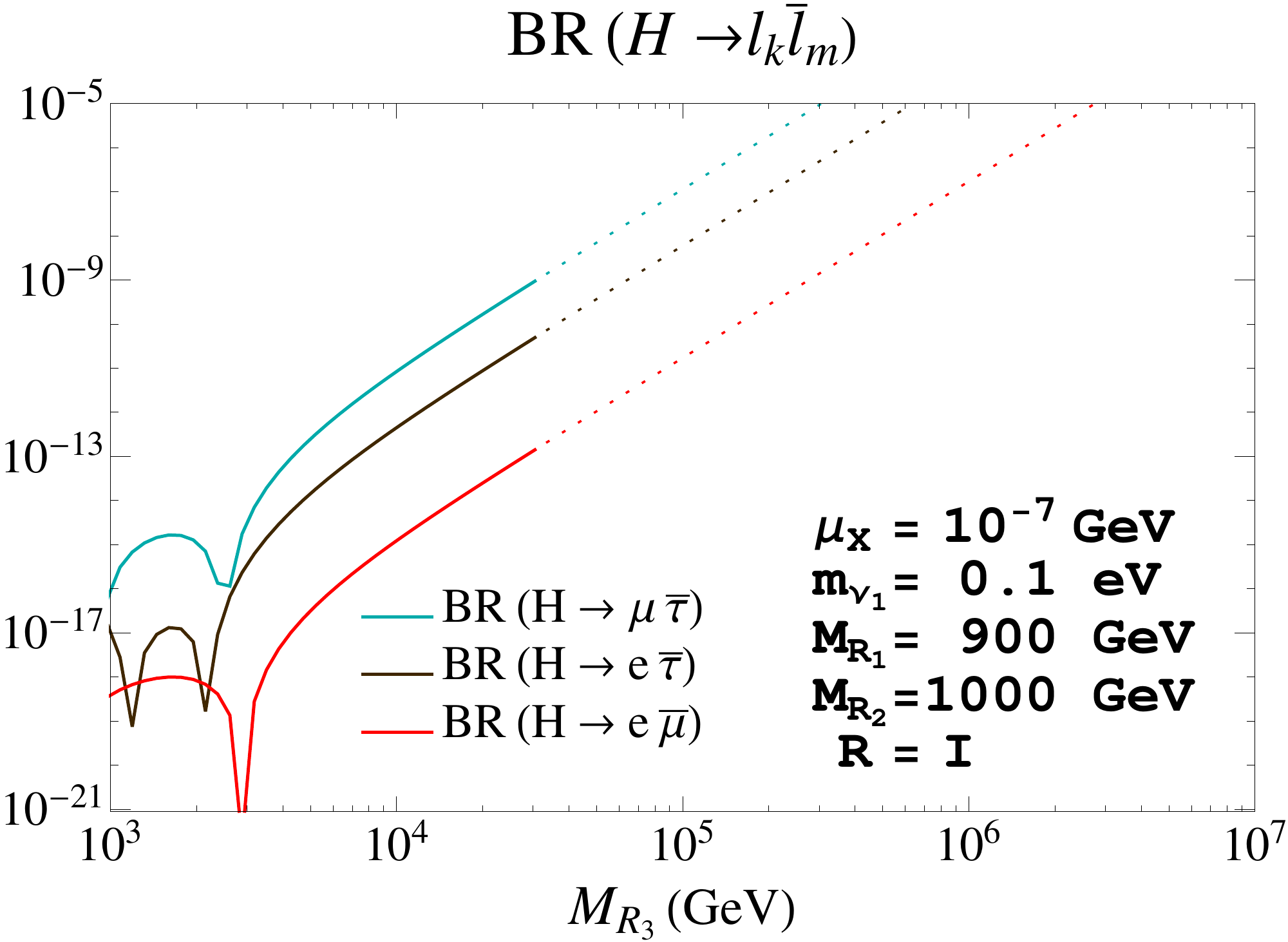} &
\includegraphics[width=0.49\textwidth]{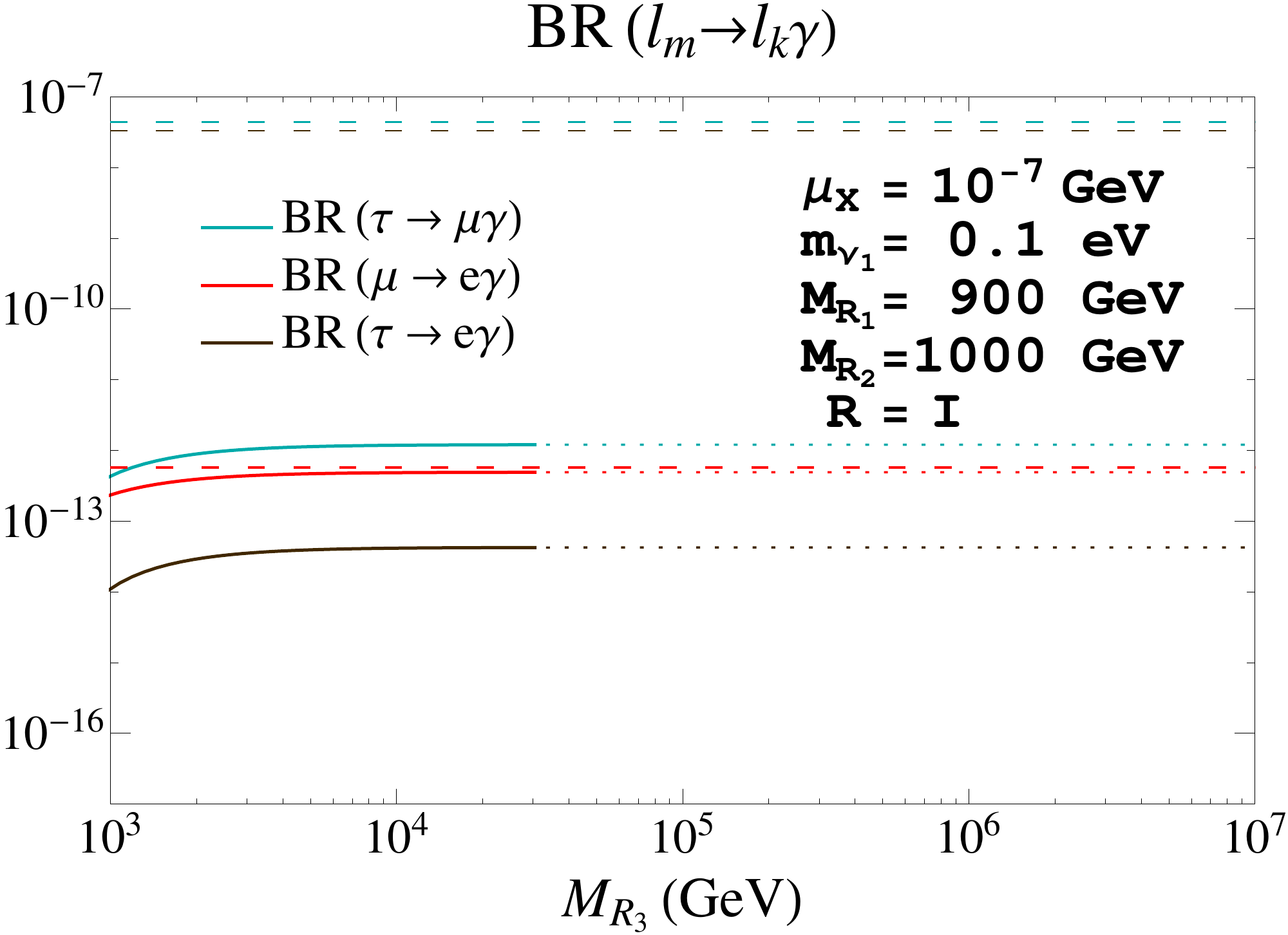}
\end{tabular}
\caption{Predictions for the LFV decay rates as functions of $M_{R_3}$ in the hierarchical heavy neutrinos case with $M_{R_1}<M_{R_2}<M_{R_3}$. Left panel:
BR$(H \to \mu \bar \tau)$ (upper blue line), BR$(H \to e \bar \tau)$ (middle dark brown line), BR$(H \to e \bar \mu)$ (lower red line). Right panel: BR$(\tau \to \mu \gamma)$
(upper blue line), BR$(\mu \to e \gamma)$ (middle red line), BR$(\tau \to e \gamma)$ (lower dark brown line). The other input parameters are set to $\mu_X= 10^{-7} \, {\rm GeV}$, 
$m_{\nu_1}=0.1 \, {\rm eV}$, $M_{R_1}= 900 \, {\rm GeV}$, $M_{R_2}= 1000 \, {\rm GeV}$, and $R=I$. The dotted lines in both panels indicate nonperturbative neutrino Yukawa couplings.
The horizontal dashed lines in the right panel are the present ($90\%$ C.L.) upper bounds on the radiative decays: BR$(\tau \to \mu \gamma)<4.4 \times 10^{-8}$~\cite{Aubert:2009ag}
(blue line), BR$(\tau \to e \gamma)<3.3 \times 10^{-8}$~\cite{Aubert:2009ag} (dark brown line), and BR$(\mu \to e \gamma)< 5.7 \times 10^{-13}$~\cite{Adam:2013mnn} (red line).
}
\label{AllHierarchical}
\end{center}
\end{figure}

The results for the LFV rates in the $M_{R_1}<M_{R_2}<M_{R_3}$ hierarchical case are shown in fig.~\ref{AllHierarchical}.
This figure shows that the behavior of the LFV rates in the hierarchical case with respect to the heaviest neutrino mass $M_{R_3}$ is very similar to the one found previously
for the degenerate case with respect to the common $M_{R}$. The BR$(H \to l_k \bar l_m)$ rates grow fast with $M_{R_3}$ at large  $M_{R_3}> 3000\, {\rm GeV}$, whereas
the BR$(l_m \to l_k \gamma)$ rates stay flat with $M_{R_3}$. Again, there are dips in the BR$(H \to l_k \bar l_m)$ rates due to the destructive interferences among the
contributing diagrams. 
We also observe in this plot that, for the chosen parameters, the size that the BR$(H \to l_k \bar l_m)$ rates can reach in this hierarchical scenario is larger than in the
previous degenerate case. For instance, BR$(H \to \mu \bar \tau)$ reaches $10^{-9}$ at 
$M_{R_3}=3\times 10^{4}\,{\rm GeV}$, to be compared with $10^{-10}$ at $M_{R}=3\times 10^{4}\,{\rm GeV}$ that we got in fig.~\ref{ALL_LFVdecays} for the degenerate case. We
have found this same behavior of enhanced LFVHD rates by approximately one order of magnitude  in the hierarchical case as compared to the degenerate case in most of the
explored parameter space regions. 

\begin{figure}[t!]
\begin{center}
\begin{tabular}{cc}
\hspace{-0.5cm}\includegraphics[width=.6\textwidth]{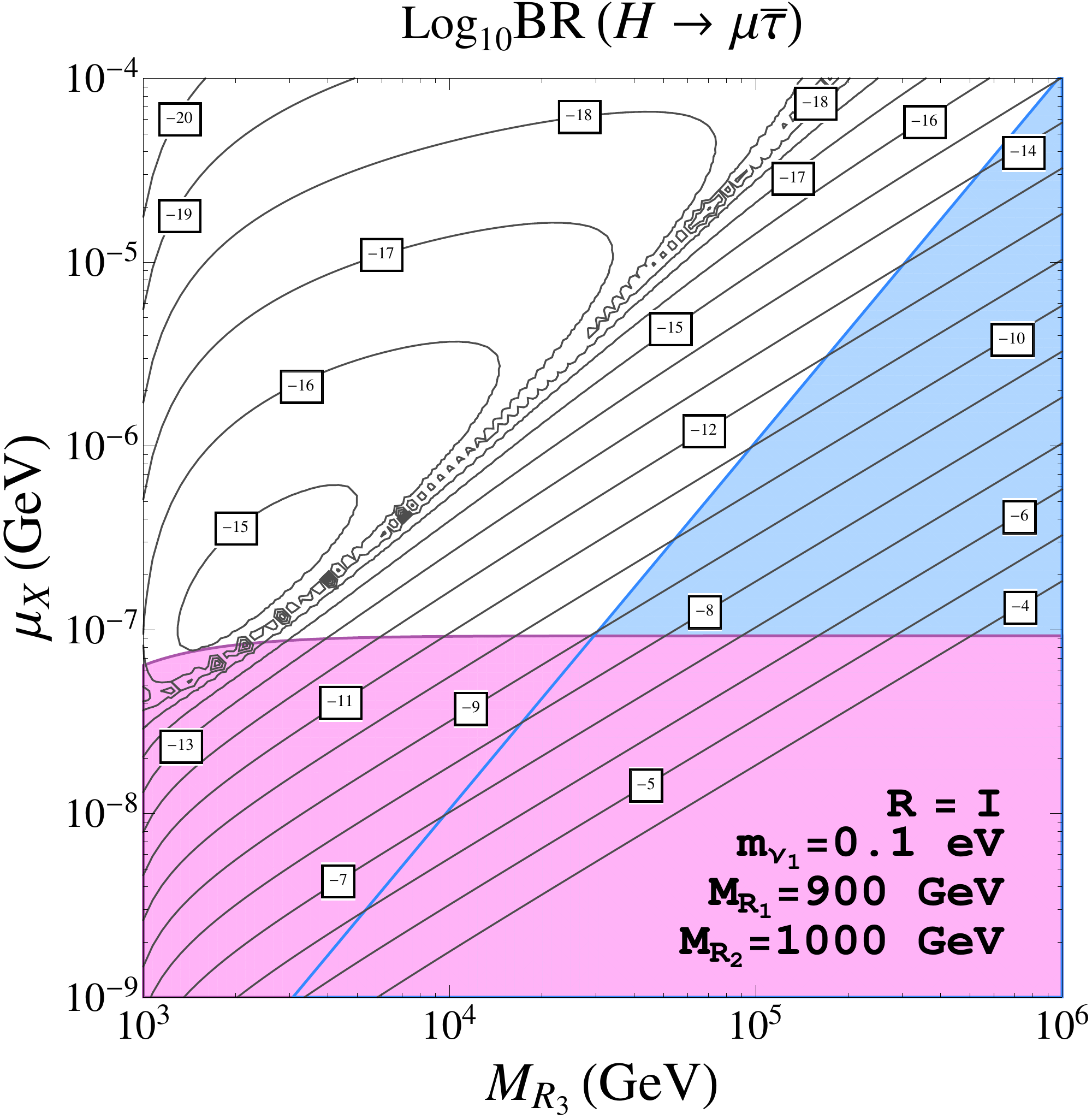} 
\end{tabular}
\caption{Contour lines of BR($H \to \mu \bar \tau$) in the $(M_{R_3},\mu_X)$ plane for $R=I$, $m_{\nu_1}=0.1\,{\rm eV}$, $M_{R_1}=900\,{\rm GeV}$, and $M_{R_2}=1000\,{\rm GeV}$. The
horizontal area in pink is disallowed by the upper bound on  BR$(\mu \to e \gamma)$. The oblique area in blue is disallowed by the perturbativity requirement of the neutrino Yukawa
couplings.}
\label{ContourPlot_MR3}
\end{center}
\end{figure}

\begin{figure}[t!]
\begin{center}
\begin{tabular}{cc}
\includegraphics[width=0.49\textwidth]{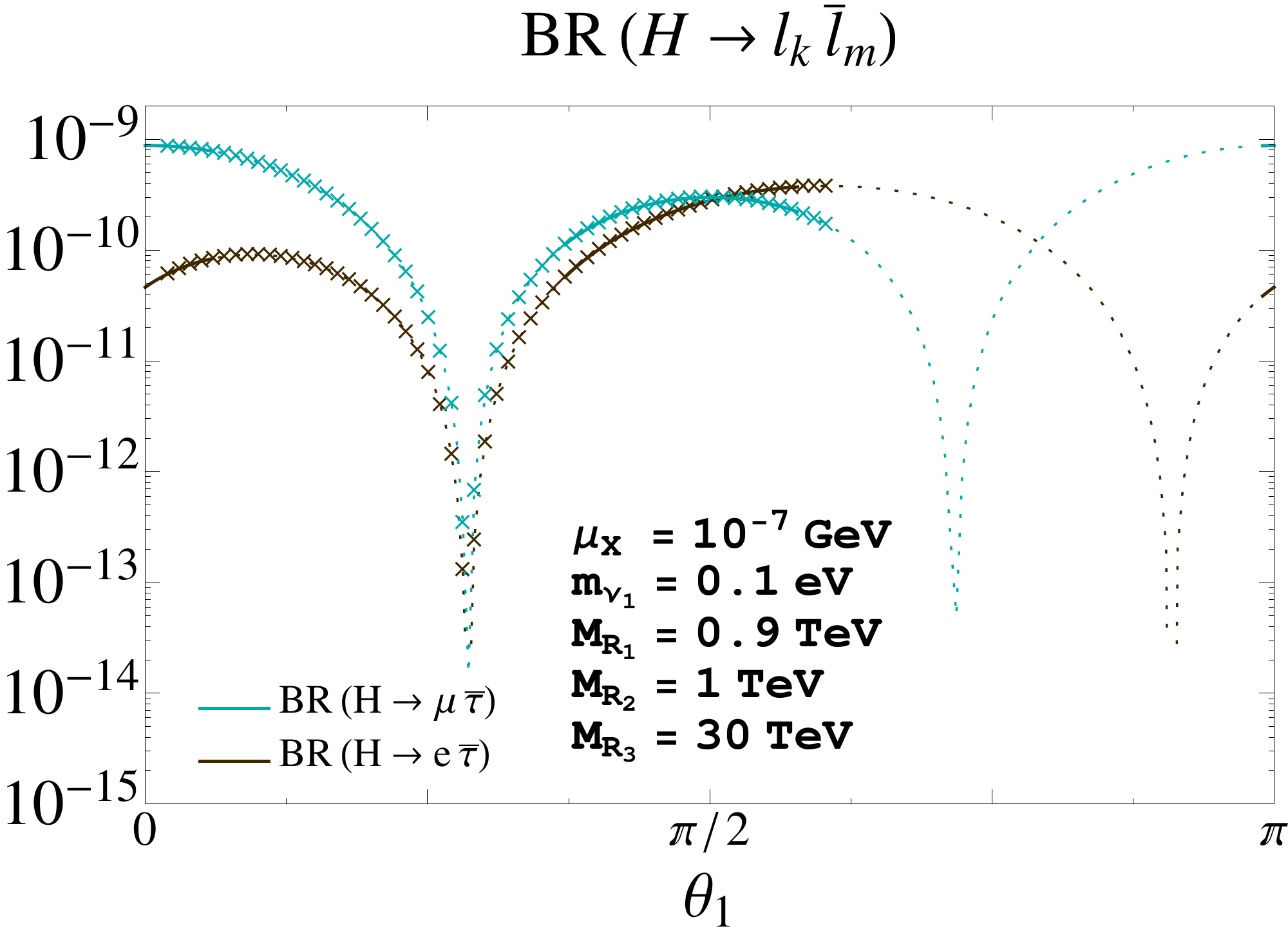} &
\includegraphics[width=0.49\textwidth]{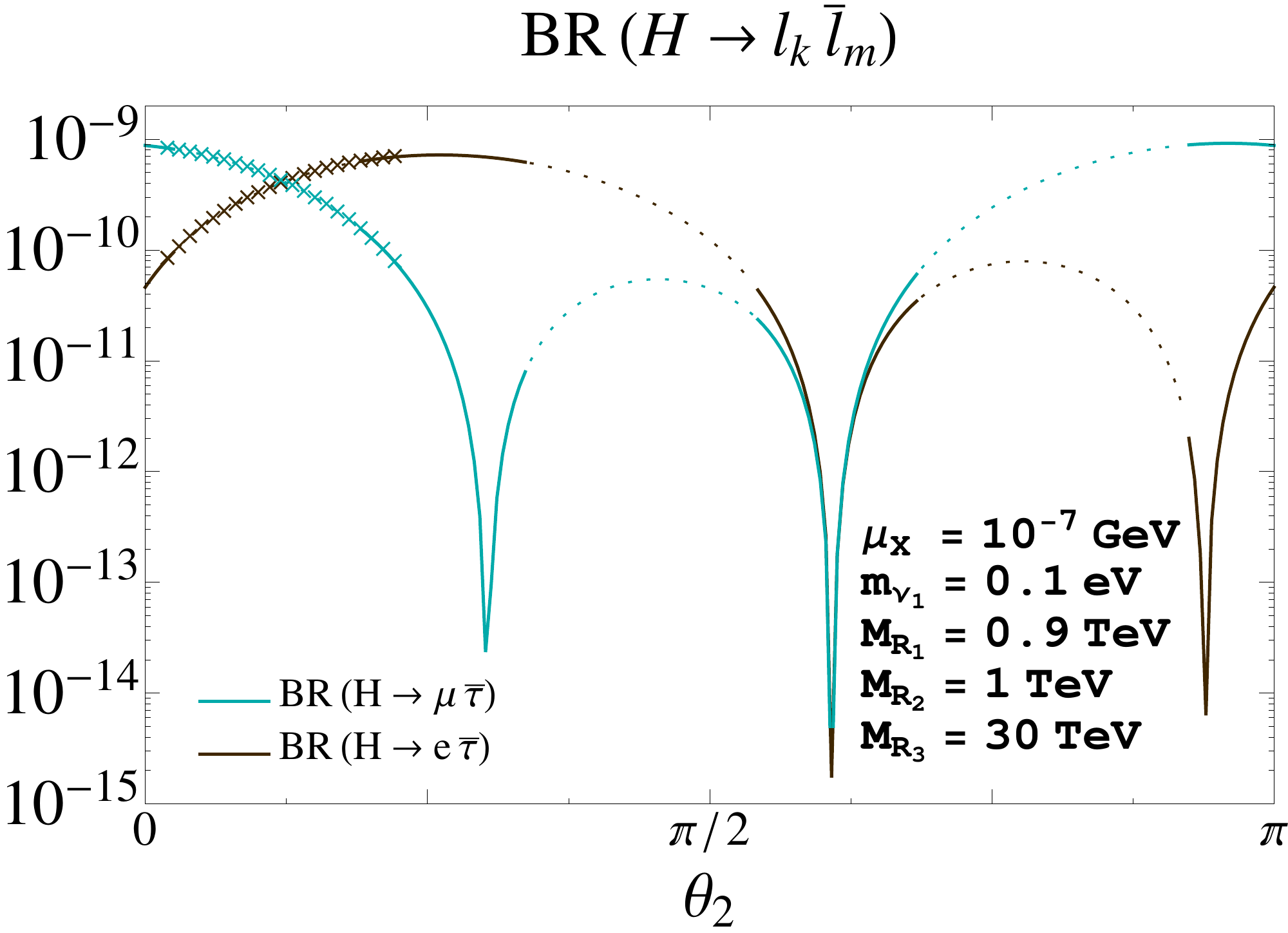}
\end{tabular}
\caption{Predictions for BR($H\to \mu\bar\tau$) (blue lines) and BR($H\to e\bar\tau$) (dark brown lines) rates as a function of real $\theta_1$ (left panel)
and $\theta_2$ (right panel). The other input parameters are set to $\mu_X=10^{-7}$ GeV, $m_{\nu_1}=0.1$ eV, $M_{R_1}=0.9$ TeV, $M_{R_2}=1$ TeV, $M_{R_3}=30$ TeV,
$\theta_2=\theta_3=0$ in the left panel and $\theta_1=\theta_3=0$ in the right panel. The dotted lines indicate nonperturbative neutrino Yukawa couplings and the crossed lines
are excluded by the present upper bound on BR($\mu\to e\gamma$). The solid lines are allowed by all the constraints.}
\label{hiertheta1and2}
\end{center}
\end{figure}

This same enhancement can also be seen in the contour plot in fig.~\ref{ContourPlot_MR3} where the maximum allowed BR$(H \to \mu \bar \tau)$ rates reach values up to about 
$10^{-9}$ for  $M_{R_1}=900\,{\rm GeV}$, $M_{R_2}=1000\,{\rm GeV}$, $M_{R_3}=3\times 10^{4}\,{\rm GeV}$, $\mu_X=10^{-7}\,{\rm GeV}$, and $R=I$.  
Finally, since in the hierarchical case, in contrast to the degenerate case, there is a dependence on the $R$ matrix even if it is real, we have also explored the behavior with
the real $\theta_{1,2,3}$ angles. We have found that for this particular hierarchy, $M_{R_1}<M_{R_2}<M_{R_3}$, there is near independence with $\theta_3$ but there is a clear
dependence with $\theta_1$ and $\theta_2$, as it is illustrated in fig.~\ref{hiertheta1and2}. These plots show that the BR$(H \to l_k \bar l_m)$ rates for $\theta_{1,2}\neq 0$ can
indeed increase or decrease with respect to the reference $R=I$ case. In particular, for  $0<\theta_{1}<\pi$ we find that BR($H \to \mu \bar \tau$) is always lower than for $R=I$, whereas
BR($H \to e \bar \tau$) can be one order of magnitude larger
than for $R=I$ if  $\theta_{1}$ is near $\pi/2$. For the case of $0<\theta_{2}<\pi$, we find again that BR($H \to \mu \bar \tau$) is always lower than for $R=I$, and
BR($H \to e \bar \tau$) can be one order of magnitude larger
than for $R=I$ if  $\theta_{2}$ is near $\pi/4$. In this latter case, it is interesting to notice that the region of $\theta_{2}$ close to $\pi/4$ where
BR($H \to e \bar \tau$) reaches the maximum value close to $10^{-9}$  is allowed by all the constraints. The results for the other decay BR($H \to e \bar \mu$) are not shown here
because they again give much smaller rates, as in the degenerate case. We have also tried other choices for the hierarchies among the three heavy masses $M_{R_{1,2,3}}$ and
we have found similar conclusions. 
\subsection{ISS scenarios with large LFV Higgs decay rates}
\label{LFVHDmax}
In this section we explore the implications on LFV Higgs decays of going beyond the simplest previous hypothesis of diagonal $\mu_X$ and $M_R$ mass matrices in the ISS model.
In particular, given the interesting possibility of  decoupling the low energy neutrino physics from the LFV physics in this ISS model, by the proper choice of the input parameters,
we will look for specific ISS scenarios with nondiagonal $\mu_X$ while keeping diagonal $M_R$ that can provide the largest LFV Higgs decay rates and at the same time be compatible
with the neutrino data and the present experimental upper bounds on the radiative decays. Here, we will focus on the case of degenerate $M_R$ and will explore only the LFV Higgs decay
channels with the largest rates, namely, $H \to \mu \bar \tau$ and $H \to e \bar \tau$. 

In order to localize the class of scenarios leading to large and allowed LFVHD rates, we first make a rough estimate of the expected maximal rates for
the $H \to \mu \bar \tau$ channel by using our approximate formula of eq.~(\ref{FIThtaumu}), which is given just in terms of the neutrino Yukawa coupling
matrix $Y_\nu$ and $M_R$. On the other hand, in order to keep the predictions for the radiative decays below their corresponding experimental upper bounds, we need to require
a maximum value for the nondiagonal $(Y_\nu Y_\nu^\dagger)_{ij}$ entries. By using our approximate formula of eq.~(\ref{approxformula}) and the present bounds
in eqs.~(\ref{MUEGmax})-(\ref{TAUMUGmax}), we get
\begin{eqnarray}
v^2(Y_\nu Y_\nu^\dagger)_{12}^{\rm max}/M_R^2 & \sim & 2.5\times10^{-5},\label{12max}\\
v^2(Y_\nu Y_\nu^\dagger)_{13}^{\rm max}/M_R^2  & \sim  & 0.015,\label{13max}\\
v^2(Y_\nu Y_\nu^\dagger)_{23}^{\rm max}/M_R^2 & \sim  & 0.017\label{23max}.
\end{eqnarray}
Then, in order to simplify our search, and given the above relative strong suppression of the 12 element, it seems reasonable to neglect it against the other
off-diagonal elements. In that case, 
by assuming $(Y_\nu Y_\nu^\dagger)_{12}\simeq 0$ we get
\begin{equation}
(Y_\nu Y_\nu^\dagger Y_\nu Y_\nu^\dagger)_{23} \simeq  (Y_\nu Y_\nu^\dagger)_{22} (Y_\nu Y_\nu^\dagger)_{23}+ (Y_\nu Y_\nu^\dagger)_{23} (Y_\nu Y_\nu^\dagger)_{33},
\end{equation} 
and the approximate formula of eq.~(\ref{FIThtaumu}) can then be rewritten as follows:
\begin{equation}
 {\rm BR}^{\rm approx}_{H\to\mu\bar\tau}=10^{-7}~\Big|\frac{v^2}{M_R^2}(Y_\nu Y_\nu^\dagger)_{23}\Big|^2~\Big| 1-5.7 \Big( (Y_\nu Y_\nu^\dagger)_{22}  +  (Y_\nu Y_\nu^\dagger)_{33}\Big) \Big|^2.
 \end{equation} 
This equation clearly shows that the maximal BR($H\to\mu\bar\tau$) rates
are obtained for the maximum allowed values of $(Y_\nu Y_\nu^\dagger)_{23}$, $(Y_\nu Y_\nu^\dagger)_{22}$, and $(Y_\nu Y_\nu^\dagger)_{33}$. Thus, before going to any
specific assumption for the $Y_\nu$ texture we can already conclude on these maximal rates, by setting the maximum allowed value
for $v^2(Y_\nu Y_\nu^\dagger)_{23}^{\rm max}/M_R^2$ to that given in eq.~(\ref{23max}) and fixing the values of $(Y_\nu Y_\nu^\dagger)_{22}$ and $(Y_\nu Y_\nu^\dagger)_{33}$ to
their maximum allowed values that are implied by our perturbativity condition in eq.~(\ref{Ymax}),
\begin{equation}
(Y_\nu Y_\nu^\dagger)_{33}^{\rm max}= (Y_\nu Y_\nu^\dagger)_{22}^{\rm max}= (Y_\nu Y_\nu^\dagger)_{11}^{\rm max}= 18 \pi ~.
\label{pert}
\end{equation}
This leads to our approximate prediction for the maximal rates:
\begin{equation}
 {\rm BR}^{\rm max}_{H\to\mu\bar\tau} \simeq 10^{-5}.
 \end{equation} 
We obtain similar conclusions for the $H\to e\bar\tau$ channel. This can be easily derived from the corresponding approximate formula that we have also checked
to work quite well in this case:
\begin{equation}\label{FIThtaue}
{\rm BR}^{\rm approx}_{H\to e\bar\tau}=10^{-7}\frac{v^4}{M_R^4}~\Big|(Y_\nu Y_\nu^\dagger)_{13}-5.7(Y_\nu Y_\nu^\dagger Y_\nu Y_\nu^\dagger)_{13}\Big|^2,
\end{equation}
leading for $(Y_\nu Y_\nu^\dagger)_{12}\simeq 0$ to
\begin{equation}
 {\rm BR}^{\rm approx}_{H\to e\bar\tau}=10^{-7}~\Big|\frac{v^2}{M_R^2}(Y_\nu Y_\nu^\dagger)_{13}\Big|^2~\Big| 1-5.7 \Big( (Y_\nu Y_\nu^\dagger)_{11}  +  (Y_\nu Y_\nu^\dagger)_{33}\Big) \Big|^2,
 \end{equation} 
 and, therefore, by using eqs.~(\ref{13max}) and (\ref{pert}) we also obtain
 \begin{equation}
 {\rm BR}^{\rm max}_{H\to e\bar\tau} \simeq 10^{-5}.
 \end{equation}  
Having such large and allowed by data LFVHD rates of the order of $10^{-5}$ for either $H\to \mu \bar\tau$ or $H\to e \bar\tau$ is clearly of great interest if the high number of Higgs events mentioned in the introduction is finally achieved.

In the rest of this section we will look for specific examples where the above 
settings can be reached. In particular, we will devote our attention to the search of particular choices of $Y_\nu$ that fulfill all the above requirements. Once some
specific inputs are provided for $Y_\nu$ and $M_R$, the proper $\mu_X$ matrix that ensures the agreement between low energy neutrino predictions and data 
can be easily obtained by solving eqs.~(\ref{Mlight})-(\ref{mnulight}), which leads to
\begin{equation}
\mu_X=M_R^T ~m_D^{-1}~ U_{\rm PMNS}^* m_\nu U_{\rm PMNS}^\dagger~ {m_D^T}^{-1} M_R
\end{equation}
with $m_D= v Y_\nu$ and $m_\nu =\mathrm{diag}(m_{\nu_1}\,, m_{\nu_2}\,, m_{\nu_3})$. 
It should be noted that for a generic $Y_\nu$ texture this $\mu_X$ will be in general nondiagonal, as announced at the beginning of this section. 

For our purpose of looking for specific examples of $Y_\nu$ maximizing the LFVHD rates and for simplicity in that search, we focus next on the case of real $Y_\nu$ where
$(Y_\nu Y_\nu^\dagger)=(Y_\nu Y_\nu^T)$, and we use a geometrical picture where the elements of the Yukawa matrix can be interpreted as the components of three vectors that we
call here $\e$, $\m$, and $\t$:
\begin{equation}\label{vectors}
Y_\nu=\left(\begin{array}{ccc} 
Y_{\nu \, 11} & Y_{\nu \,12} & Y_{\nu\,13}\\Y_{\nu\,21}&Y_{\nu\,22}&Y_{\nu\,23}\\ Y_{\nu\,31}&Y_{\nu \,32}&Y_{\nu \,33}\end{array}\right)\equiv
\left(\begin{array}{c} \e\\\m\\ \t \end{array}\right).
\end{equation}
Then the relevant matrix for our LFV observables can be written as
\begin{equation}
YY^T=\left(\begin{array}{ccc}
 |\e|^2 & \e\cdot\m &\e\cdot\t \\ \m\cdot\e & |\m|^2 &\m\cdot\t \\ \t\cdot\e &\t\cdot\m& |\t|^2
 \end{array}\right),
 \end{equation} 
and consequently it can be completely determined by  
setting six parameters: the modulus of the three vectors $(|\e|,|\m|,|\t|)$ and the three angles $(\tme,\tte,\ttm)$  defining their relative orientations.
It should be noticed, however, that a real $3 \times 3$ Yukawa matrix should contain nine parameters. The missing three parameters can be understood in terms of an additional
rotation $O$ of the 3 vectors, which does not change their relative angles, and therefore it has no physical consequences for our observables. Thus, one can write, generically,
the neutrino Yukawa matrix as a product of two matrices 
$A$ and $O$, with $OO^T=O^TO=1$:  
\begin{equation}
Y_\nu \equiv A\cdot O ,
\end{equation}
\begin{equation}
Y_\nu Y_\nu^T=A A^T.
\end{equation}
 
We can use then this freedom to choose the two orthogonal vectors $(\e,\m)$ in two of the axes, for instance $\e$ in the $X$ axis and $\m$ in the $Y$ axis, so that we can write
\begin{equation}\label{A}
A=\left(\begin{array}{ccc}
e& 0& 0\\ 0&\mu&0\\ \tau\cte&\tau\ctm&\tau\sqrt{1-\cte^2-\ctm^2}
\end{array}\right),
 \end{equation}
with $|\e| \equiv e$, $|\m| \equiv \mu$, $|\t| \equiv \tau$, $\cte\equiv\cos\tte$, and $\ctm\equiv\cos\ttm$.
Then in our simple geometrical parametrization of the Yukawa matrix we get
\begin{equation}
Y_\nu Y_\nu^T=A A^T=\left(\begin{array}{ccc}
 e^2 & 0 &e\tau\cte \\ 0 & \mu^2 &\mu\tau\ctm \\ e\tau\cte &\mu\tau\ctm& \tau^2
 \end{array}\right)
\end{equation}
which shows explicitly our requirement of $(Y_\nu Y_\nu^T)_{12}=0$ and whose simple form helps in the choice of the textures maximizing the LFVHD rates. For instance,
it is obvious that by choosing parallel or antiparallel $\t$ and $\m$ vectors, i.e., $\ctm=\pm 1$  we will get maximal BR$(H \to \mu \bar \tau)$,  whereas, by choosing
parallel or antiparallel $\t$ and $\e$ vectors, i.e., $\cte=\pm 1$  we will get maximal BR$(H \to e \bar \tau)$. We also see that we will not be able to get maximal
rates for both channels simultaneously, since the imposed orthogonality of $\e$ and $\m$ implies some correlations among the LFV in the $e-\tau$ and $\mu-\tau$ channels. Thus,
for a given input $\ttm$, the maximum   LFV rates in the $e-\tau$ channel will occur at the correlated value $\tte= \pi/2- \ttm$, and vice versa.  As a consequence, the
maximum in BR$(H \to \mu \bar \tau)$ implies a minimum in   BR$(H \to e \bar \tau)$, and a maximum in BR$(H \to e \bar \tau)$ implies a minimum in BR$(H \to \mu \bar \tau)$. We 
find this result an interesting feature of this kind of texture. 

Finally, we provide some illustrative examples with large LFVHD rates. All of them fulfil $(Y_\nu Y_\nu^T)_{12}=0$ and $(Y_\nu Y_\nu^T Y_\nu Y_\nu^T)_{12}=0$, therefore ensuring
the practically vanishing LFV in the $\mu-e$ sector, i.e., leading all to BR$(\mu \to e \gamma)\sim 0$ and BR$(H \to e \bar \mu)\sim 0$.

(1) Examples with large LFV $\mu-\tau$:

The following three textures,  $Y_{\tau \mu}^{(1)}$, $Y_{\tau \mu}^{(2)}$, and  $Y_{\tau \mu}^{(3)}$, provide large LFV in the $\mu-\tau$ sector, and practically
vanishing LFV in the $e-\tau$ sector, since they all have $\e\cdot\t=0$: 
\begin{equation}
Y_{\tau \mu}^{(1)}=\sqrt{6\pi}\left(\begin{array}{ccc}
0&1&-1\\0.9&1&1\\1&1&1
\end{array}\right)~,~
Y_{\tau \mu}^{(2)}=\sqrt{6\pi}\left(\begin{array}{ccc}
0&1&1\\1&1&-1\\-1&1&-1
\end{array}\right)~,~
Y_{\tau \mu}^{(3)}=\sqrt{6\pi}\left(\begin{array}{ccc}
0&-1&1\\-1&1&1\\0.8&0.5&0.5
\end{array}\right).
\label{Ytmmax}
\end{equation}
These textures $Y_{\tau \mu}^{(1,2,3)}$ can be obtained by choosing $A_{\tau \mu}^{(1,2,3)}$ matrices like the $A$ matrix in eq.~(\ref{A}) with
$e^{(1,2,3)}=(\sqrt{12 \pi}, \sqrt{12 \pi},\sqrt{12 \pi})$, $\mu^{(1,2,3)}=(\sqrt{17.4 \pi}, \sqrt{18 \pi},\sqrt{18 \pi})$,
$\tau^{(1,2,3)}=(\sqrt{18 \pi}, \sqrt{18 \pi},\sqrt{6.4 \pi})$, and $\ctm^{(1,2,3)}=(0.98,0.33,0.025)$, respectively; and then applying the
corresponding rotation $O_{\tau \mu}=(A^{-1}_{\tau \mu} )Y_{\tau \mu}$.

(2) Examples with large LFV $e-\tau$:

The following three textures,  $Y_{\tau e}^{(1)}$, $Y_{\tau e}^{(2)}$, and  $Y_{\tau e}^{(3)}$, provide large LFV in the $e-\tau$ sector, and practically
vanishing LFV in the $\mu-\tau$ sector, since they all have $\t\cdot\m=0$: 
\begin{equation}
Y_{\tau e}^{(1)}=\sqrt{6\pi}\left(\begin{array}{ccc}
0.9&1&1\\0&1&-1\\1&1&1
\end{array}\right)~,~
Y_{\tau e}^{(2)}=\sqrt{6\pi}\left(\begin{array}{ccc}
1&1&-1\\0&1&1\\-1&1&-1
\end{array}\right)~,~
Y_{\tau e}^{(3)}=\sqrt{6\pi}\left(\begin{array}{ccc}
-1&1&1\\0&-1&1\\0.8&0.5&0.5
\end{array}\right)
\label{Ytemax}
\end{equation}
These textures $Y_{\tau e}^{(1,2,3)}$  can be obtained by choosing $A_{\tau e}^{(1,2,3)}$ matrices like the $A$ matrix in eq.~(\ref{A})
with $e^{(1,2,3)}=(\sqrt{17.4 \pi}, \sqrt{18 \pi},\sqrt{18 \pi})$, $\mu^{(1,2,3)}=(\sqrt{12 \pi}, \sqrt{12 \pi},\sqrt{12 \pi})$,
$\tau^{(1,2,3)}=(\sqrt{18 \pi}, \sqrt{18 \pi},\sqrt{6.4 \pi})$, and $\cte^{(1,2,3)}=(0.98,0.33,0.025)$, respectively; and then applying the corresponding
rotation $O_{\tau e}=A^{-1}_{\tau e} Y_{\tau e}$.
 
\begin{figure}[t!]
\begin{center}
\begin{tabular}{cc}
\includegraphics[width=0.49\textwidth]{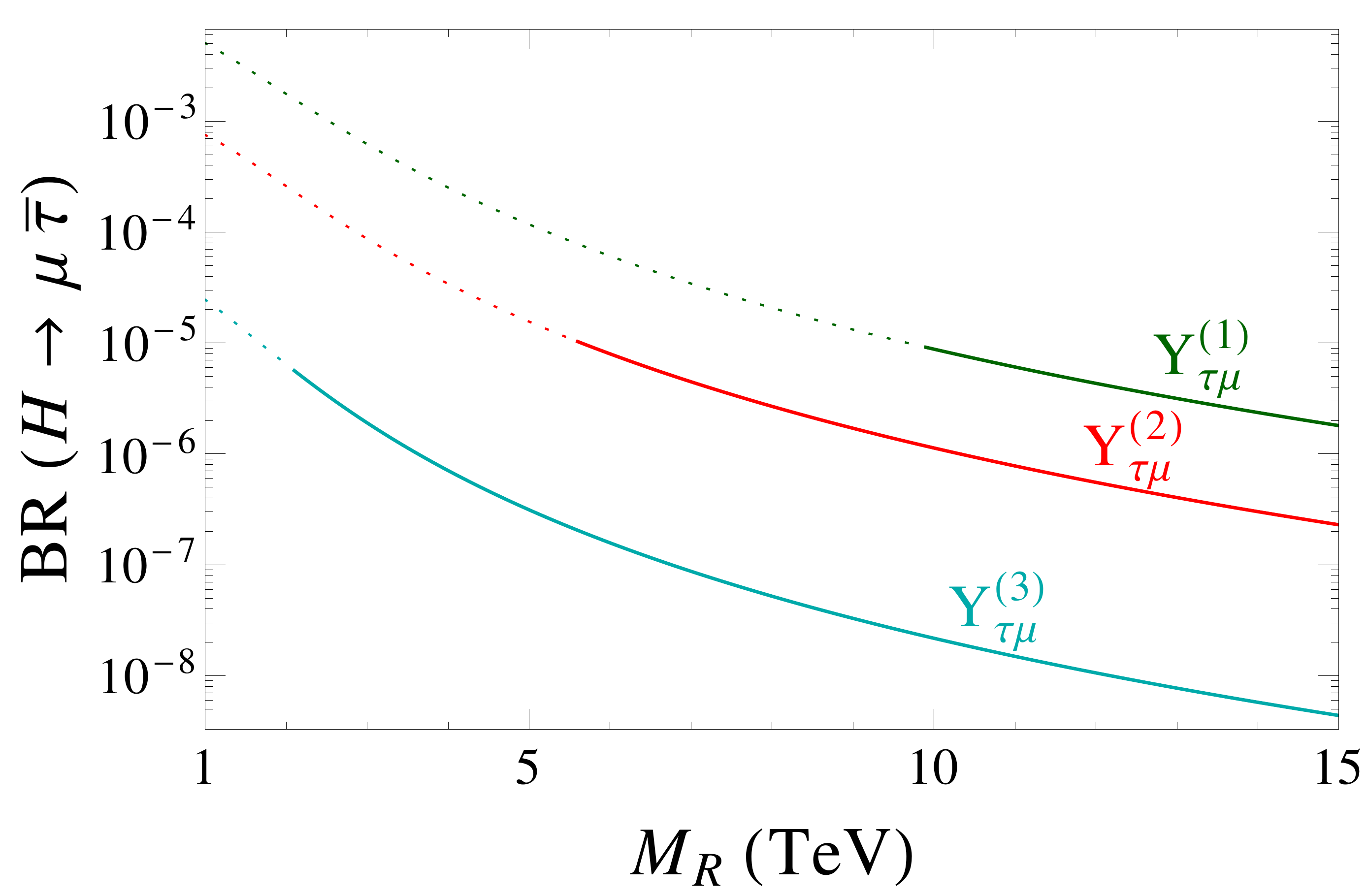} &
\includegraphics[width=0.49\textwidth]{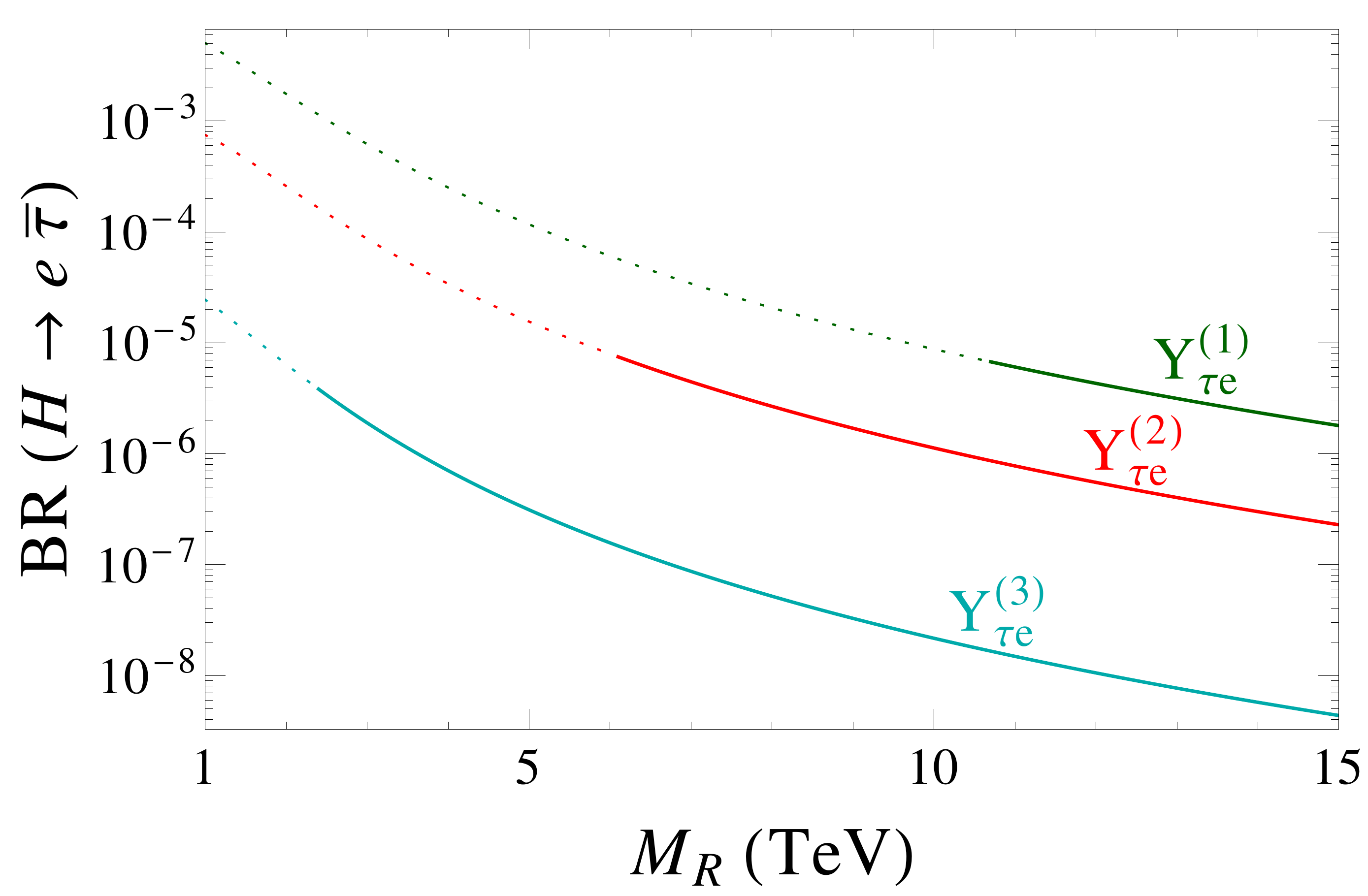}
\end{tabular}
\caption{Examples in the ISS with large LFVHD rates obtained using the full one-loop formulas. Left panel:  BR$(H \to \mu \bar \tau)$ versus $M_R$ for
$Y_{\tau \mu}^{(1)}$ (upper green line), $Y_{\tau \mu}^{(2)}$ (middle red line) and  $Y_{\tau \mu}^{(3)}$ (lower blue line) given in eq.~(\ref{Ytmmax}). Dotted lines
indicate disallowed input values leading to BR$(\tau \to \mu \gamma)$ above the present experimental bound in eq.(\ref{TAUMUGmax}). Right panel:  BR$(H \to e \bar \tau)$ versus
$M_R$ for $Y_{\tau e}^{(1)}$ (upper green line), $Y_{\tau e}^{(2)}$ (middle red line), and  $Y_{\tau e}^{(3)}$ (lower blue line) given in eq.~(\ref{Ytemax}). Dotted lines indicate
disallowed input values leading to BR$(\tau \to e \gamma)$ above the present experimental bound in eq.~(\ref{TAUEGmax}). Solid lines indicate predictions allowed by all the constraints.}
\label{maxLFVHD}
\end{center}
\end{figure}

We present our predictions for the LFVHD rates in our above selected examples  in fig.~\ref{maxLFVHD} as a function of the degenerate right-handed neutrino mass $M_R$. For
these predictions we have used the full one-loop formulas. We have also checked that the approximate formulas in eqs.~(\ref{FIThtaumu}) and (\ref{FIThtaue}) give a quite good
estimate of these BRs in the large $M_R$ region, with deviations with respect to the full result smaller than $10\%$ for $M_R> 6 \,{\rm TeV}$.  The main conclusion from these plots
is that with these specific Yukawa textures one can indeed reach large LFVHD rates of the order of $10^{-5}$ and still be compatible with all the bounds from radiative
decays. The textures $Y_{\tau \mu}$ ($Y_{\tau e}$) corresponding to lower $\ctm$ ($\cte$) allow for lower $M_R$ values and vice versa. Thus, $Y_{\tau \mu}^{(1)}$ ($Y_{\tau e}^{(1)}$) leads the maximum allowed BR$(H \to \mu \bar \tau)$ (BR$(H \to e \bar \tau)$) rates for $M_R$ around
10 TeV (11 TeV), $Y_{\tau \mu}^{(2)}$ ($Y_{\tau e}^{(2)}$) around 5.5 TeV (6 TeV), and 
$Y_{\tau \mu}^{(3)}$ ($Y_{\tau e}^{(3)}$) around 2 TeV (2.5 TeV). 
 
The above textures are just some selected examples, among many possibilities, but the important feature is that they will all provide maximum allowed rates of around $10^{-5}$. 
We have also checked that by selecting examples with hierarchical $M_{R_1}$, $M_{R_2}$, $M_{R_3}$ masses we do not obtain larger maximum allowed rates. Thus, our
conclusion is quite generic for the maximum allowed LFVHD rates in the ISS models. The other generic feature that is worth mentioning is that, given the correlated rates
found between   
${\rm BR}_{\rm max}(H \to \mu \bar \tau)$ and  ${\rm BR}_{\rm max}(\tau \to \mu \gamma)$ [similarly, between   
${\rm BR}_{\rm max}(H \to e \bar \tau)$ and  ${\rm BR}_{\rm max}(\tau \to e  \gamma)$], if an improved future upper experimental bound
on ${\rm BR}(\tau \to \mu \gamma)$ [similarly, on ${\rm BR}(\tau \to e  \gamma)$] is provided, this will be intermediately translated into a smaller maximal allowed value
for ${\rm BR}(H \to \mu \bar \tau)$ [similarly, for ${\rm BR}(H \to e \bar \tau)$].

\section{Conclusions}
In this paper we have studied the LFV Higgs decays  $H \to l_k\bar l_m$ within the context of the inverse seesaw model where three additional pairs (one pair per generation) of
massive right-handed singlet neutrinos are added to the standard model particle content. 
We have presented a full one-loop computation of the BR($H \to l_k\bar l_m$) rates for the three possible channels, $l_k\bar l_m=\mu \bar \tau, e \bar \tau, e \bar \mu$, and have
analyzed in full detail the predictions as functions of the various relevant ISS parameters. The most relevant parameters for LFV have been found to be $M_{R}$ and $Y_\nu$.
In addition, we have required that the input parameters of this ISS model be compatible with the present neutrino data and other constraints, like perturbativity of the neutrino
Yukawa couplings and the present bounds for the three radiative decays  $\mu\to e\gamma$, $\tau\to e\gamma$, and $\tau\to\mu\gamma$.  To take control on this last requirement, we
have studied along this paper in parallel to the LFV Higgs decays the correlated one-loop predictions for the radiative decays, $l_m \to l_k \gamma$, within this same ISS context. 
We have explored the ISS parameter space and consider both kinds of scenarios for the right-handed neutrinos, with either degenerate or hierarchical masses.  First, we have
considered the simplest case of diagonal $M_R$ and $\mu_X$ matrices. In this case, we conclude that the largest maximum LFV Higgs decay rates within the ISS that are allowed
by all the constraints are for BR($H \to e \bar \tau$) and BR($H \to \mu \bar \tau$) and reach at most $10^{-10}$ for the degenerate heavy neutrino case and $10^{-9}$ for the
hierarchical case. Second, we have explored more general ISS scenarios with nondiagonal $\mu_X$ matrices that we have found more promising for LFVHD searches. These can also
accommodate successfully the low energy neutrino data, and be compatible with the present bounds on the radiative decays and with the perturbativity bounds on the neutrino Yukawa
couplings. We have demonstrated that in this kind of ISS scenarios there are solutions with much larger allowed LFVHD rates than in the previous cases, 
leading to maximal allowed rates of around $10^{-5}$ for either BR($H \to \mu \bar \tau$) or BR($H \to e \bar \tau$). Assuming in addition $CP$ conservation in these scenarios, the final LFVHD rates should be multiplied by a factor of 2 if the $CP$ conjugate channels BR($H \to \tau \bar \mu$) and BR($H \to \tau \bar e$) are also considered. Finally, we have also provided a few particular
examples where the predicted rates with the full one-loop formulas indeed give such a large LFVHD rate of $\sim 10^{-5}$, for values of $M_R$ in the interval (1 TeV, 10 TeV). We certainly
find these LFVHD rates and $M_R$ values interesting, given the expected extremely high statistics of up to hundreds of millions of Higgs bosons that will be produced at the future colliders, allowing for searches of rare Higgs decays, and the potential of LHC to explore new particles at the TeV region.  
 
\label{conclusions}

\section*{Acknowledgements}

This work is supported by the European Union Grant No. FP7 ITN
INVISIBLES (Marie Curie Actions, Grant No. PITN- GA-2011- 289442), by the CICYT through Grant No. FPA2012-31880,  
by the Spanish Consolider-Ingenio 2010 Programme CPAN (Grant No. CSD2007-00042), 
and by the Spanish MINECO's ``Centro de Excelencia Severo Ochoa'' Programme under Grant No. SEV-2012-0249.
E.~A. is financially supported by the Spanish DGIID-DGA Grant No. 2013-E24/2 and the Spanish MICINN Grants No. FPA2012-35453 and No. CPAN-CSD2007-00042.
X.~M. is supported through the FPU Grant No. AP-2012-6708.

\newpage
\appendix 

\vspace{1cm}
{\Large{\bf Appendix Analytical expressions of the form factors}}

\vspace{0.5cm}
\label{App:A}
For completeness, we collect here the analytical results for the LFV Higgs decay form factors  in the Feynman 't Hooft gauge and expressed in the physical basis. These
formulas are taken from ref.\cite{Arganda:2004bz}.
\begin{eqnarray*}
&F_L^{(1)}& = \frac{g^2}{4 m_W^3} \frac{1}{16 \pi^2} B_{l_k n_i} B_{l_m n_j}^* \left\{m_{l_k} m_{n_j} \left[(m_{n_i} + m_{n_j}) \mbox{Re}\left(C_{n_i n_j}\right) + i(m_{n_j} - m_{n_i}) \mbox{Im}\left(C_{n_i n_j}\right) \right] \tilde{C}_0 \right. \\
&&+ (C_{12} - C_{11}) \left[(m_{n_i} + m_{n_j}) \mbox{Re}\left(C_{n_i n_j}\right) \left( -m_{l_k}^3 m_{n_j} - m_{n_i} m_{l_k} m_{l_m}^2 + m_{n_i} m_{n_j}^2 m_{l_k} + m_{n_i}^2 m_{n_j} m_{l_k} \right) \right. \\
&&+ \left. \left. i(m_{n_j} - m_{n_i}) \mbox{Im}\left(C_{n_i n_j}\right) \left( -m_{l_k}^3 m_{n_j} + m_{n_i} m_{l_k} m_{l_m}^2 - m_{n_i} m_{n_j}^2 m_{l_k} + m_{n_i}^2 m_{n_j} m_{l_k} \right) \right] \right\}\,,\\
&F_R^{(1)}& = \frac{g^2}{4 m_W^3} \frac{1}{16 \pi^2} B_{l_k n_i} B_{l_m n_j}^* \left\{ m_{n_i} m_{l_m} \left[(m_{n_i} + m_{n_j}) \mbox{Re}\left(C_{n_i n_j}\right) - i(m_{n_j} - m_{n_i}) \mbox{Im}\left(C_{n_i n_j}\right) \right] \tilde{C}_0 \right. \\
&&+ C_{12} \left[(m_{n_i} + m_{n_j}) \mbox{Re}\left(C_{n_i n_j}\right) \left( m_{l_m}^3 m_{n_i} - m_{n_i} m_{n_j}^2 m_{l_m} - m_{n_i}^2 m_{n_j} m_{l_m} + m_{n_j} m_{l_k}^2 m_{l_m} \right) \right. \\
&&+ \left. \left. i(m_{n_j} - m_{n_i}) \mbox{Im}\left(C_{n_i n_j}\right) \left( -m_{l_m}^3 m_{n_i} + m_{n_i} m_{n_j}^2 m_{l_m} - m_{n_i}^2 m_{n_j} m_{l_m} + m_{n_j} m_{l_k}^2 m_{l_m} \right) \right] \right\}\,,
\end{eqnarray*}
where $C_{11, 12} = C_{11, 12}(m_{l_k}^2, m_H^2, m_W^2, m_{n_i}^2, m_{n_j}^2)$ and $\tilde{C}_0 = \tilde{C}_0(m_{l_k}^2, m_H^2, m_{W }^2, m_{n_i}^2, m_{n_j}^2)$.
\begin{eqnarray*}
&F_L^{(2)}& = \frac{g^2}{2 m_W} \frac{1}{16 \pi^2} B_{l_k n_i} B_{l_m n_j}^* m_{l_k} \left\{-m_{n_j} \left[(m_{n_i} + m_{n_j}) \mbox{Re}\left(C_{n_i n_j}\right) + i(m_{n_j} - m_{n_i}) \mbox{Im}\left(C_{n_i n_j}\right) \right] C_0 \right. \\
&&+ \left. (C_{12} - C_{11}) \left[(m_{n_i} + m_{n_j})^2 \, \mbox{Re}\left(C_{n_i n_j}\right) + i(m_{n_j} - m_{n_i})^2 \, \mbox{Im}\left(C_{n_i n_j}\right) \right] \right\}\,,\\
&F_R^{(2)}& = -\frac{g^2}{2 m_W} \frac{1}{16 \pi^2} B_{l_k n_i} B_{l_m n_j}^* m_{l_m} \left\{m_{n_i} \left[(m_{n_i} + m_{n_j}) \mbox{Re}\left(C_{n_i n_j}\right) - i(m_{n_j} - m_{n_i}) \mbox{Im}\left(C_{n_i n_j}\right) \right] C_0 \right. \\
&&+ \left. C_{12} \left[(m_{n_i} + m_{n_j})^2 \, \mbox{Re}\left(C_{n_i n_j}\right)+ i(m_{n_j} - m_{n_i})^2 \, \mbox{Im}\left(C_{n_i n_j}\right) \right] \right\}\,,
\end{eqnarray*}
where $C_{0, 11, 12} = C_{0, 11, 12}(m_{l_k}^2, m_H^2, m_W^2, m_{n_i}^2, m_{n_j}^2)$.
\begin{eqnarray*}
&F_L^{(3)}& = \frac{g^2}{16 \pi^2} B_{l_k n_i} B_{l_m n_i}^* m_{l_k} m_W \left(C_{11} - C_{12}\right)\,,\\
&F_R^{(3)}& =  \frac{g^2}{16 \pi^2} B_{l_k n_i} B_{l_m n_i}^* m_{l_m} m_W C_{12}\,,
\end{eqnarray*}
where $C_{11, 12} = C_{11, 12}(m_{l_k}^2, m_H^2, m_{n_i}^2, m_W^2, m_W^2)$.
\begin{eqnarray*}
&F_L^{(4)}& = -\frac{g^2}{4 m_W} \frac{1}{16 \pi^2} B_{l_k n_i} B_{l_m n_i}^* m_{l_k} \left\{ m_{l_m}^2 (C_{12} - 2 C_{11}) + m_{n_i}^2 (C_{11} - C_{12}) - m_{n_i}^2 C_0 \right\}\,, \\
&F_R^{(4)}& = -\frac{g^2}{4 m_W} \frac{1}{16 \pi^2} B_{l_k n_i} B_{l_m n_i}^* m_{l_m} \left\{ \tilde{C}_0 + 2 m_{l_m}^2 C_{11} + m_{n_i}^2 C_{12} + (m_{l_k}^2 - 2 m_H^2) (C_{11} - C_{12}) + 2 m_{n_i}^2 C_0 \right\}\,,
\end{eqnarray*}
where $C_{0, 11, 12} = C_{0, 11, 12}(m_{l_k}^2, m_H^2, m_{n_i}^2, m_W^2, m_W^2)$ and $\tilde{C}_0 = \tilde{C}_0(m_{l_k}^2, m_H^2, m_{n_i}^2, m_W^2, m_W^2)$.
\begin{eqnarray*}
&F_L^{(5)}& = -\frac{g^2}{4 m_W} \frac{1}{16 \pi^2} B_{l_k n_i} B_{l_m n_i}^* m_{l_k} \left\{\tilde{C}_0 + 2 m_{n_i}^2 C_0 + (m_{n_i}^2 + 2 m_{l_k}^2) C_{11} + (m_{l_m}^2 - m_{n_i}^2 - 2 m_H^2) C_{12} \right\} \,,\\
&F_R^{(5)}& = \frac{g^2}{4 m_W} \frac{1}{16 \pi^2} B_{l_k n_i} B_{l_m n_i}^* m_{l_m} \left\{m_{n_i}^2 C_0 + m_{l_k}^2 C_{11} + (m_{l_k}^2 - m_{n_i}^2) C_{12} \right\}\,,
\end{eqnarray*}
where $C_{0, 11, 12} = C_{0, 11, 12}(m_{l_k}^2, m_H^2, m_{n_i}^2, m_W^2, m_W^2)$ and $\tilde{C}_0 = \tilde{C}_0(m_{l_k}^2, m_H^2, m_{n_i}^2, m_W^2, m_W^2)$.
\begin{eqnarray*}
&F_L^{(6)}& = \frac{g^2}{4 m_W^3} \frac{1}{16 \pi^2} B_{l_k n_i} B_{l_m n_i}^* m_{l_k} m_H^2 \left\{m_{n_i}^2 (C_0 + C_{11}) + (m_{l_m}^2 - m_{n_i}^2) C_{12} \right\}\,, \\
&F_R^{(6)}& = \frac{g^2}{4 m_W^3} \frac{1}{16 \pi^2} B_{l_k n_i} B_{l_m n_i}^* m_{l_m} m_H^2 \left\{m_{n_i}^2 (C_0 + C_{12}) + m_{l_k}^2 (C_{11} - C_{12})  \right\}\,,
\end{eqnarray*}
where $C_{0, 11, 12} = C_{0, 11, 12}(m_{l_k}^2, m_H^2, m_{n_i}^2, m_W^2, m_W^2)$.
\begin{eqnarray*}
&F_L^{(7)}& = \frac{g^2}{2 m_W} \frac{1}{16 \pi^2} B_{l_k n_i} B_{l_m n_i}^* \frac{m_{l_m}^2 m_{l_k}}{m_{l_k}^2 - m_{l_m}^2} B_1 \,,\\
&F_R^{(7)}& = \frac{g^2}{2 m_W} \frac{1}{16 \pi^2} B_{l_k n_i} B_{l_m n_i}^* \frac{m_{l_k}^2 m_{l_m}}{m_{l_k}^2 - m_{l_m}^2} B_1\,, \\
&F_L^{(8)}& = \frac{g^2}{4 m_W^3} \frac{1}{16 \pi^2} B_{l_k n_i} B_{l_m n_i}^* \frac{m_{l_k}}{m_{l_k}^2 - m_{l_m}^2} \left\{ m_{l_m}^2 (m_{l_k}^2 + m_{n_i}^2) B_1 + 2 m_{n_i}^2 m_{l_m}^2 B_0 \right\} \,,\\
&F_R^{(8)}& = \frac{g^2}{4 m_W^3} \frac{1}{16 \pi^2} B_{l_k n_i} B_{l_m n_i}^* \frac{m_{l_m}}{m_{l_k}^2 - m_{l_m}^2} \left\{ m_{l_k}^2 (m_{l_m}^2 + m_{n_i}^2) B_1 + m_{n_i}^2 (m_{l_k}^2 + m_{l_m}^2) B_0 \right\}\,,
\end{eqnarray*}
where $B_{0, 1} = B_{0, 1}(m_{l_k}^2, m_{n_i}^2, m_W^2)$.
\begin{eqnarray*}
&F_L^{(9)}& = \frac{g^2}{2 m_W} \frac{1}{16 \pi^2} B_{l_k n_i} B_{l_m n_i}^* \frac{m_{l_m}^2 m_{l_k}}{m_{l_m}^2 - m_{l_k}^2} B_1\,, \\
&F_R^{(9)}& = \frac{g^2}{2 m_W} \frac{1}{16 \pi^2} B_{l_k n_i} B_{l_m n_i}^* \frac{m_{l_k}^2 m_{l_m}}{m_{l_m}^2 - m_{l_k}^2} B_1\,, \\
&F_L^{(10)}& = \frac{g^2}{4 m_W^3} \frac{1}{16 \pi^2} B_{l_k n_i} B_{l_m n_i}^* \frac{m_{l_k}}{m_{l_m}^2 - m_{l_k}^2} \left\{ m_{l_m}^2 (m_{l_k}^2 + m_{n_i}^2) B_1 + m_{n_i}^2 (m_{l_k}^2 + m_{l_m}^2) B_0 \right\}\,, \\
&F_R^{(10)}& = \frac{g^2}{4 m_W^3} \frac{1}{16 \pi^2} B_{l_k n_i} B_{l_m n_i}^* \frac{m_{l_m}}{m_{l_m}^2 - m_{l_k}^2} \left\{ m_{l_k}^2 (m_{l_m}^2 + m_{n_i}^2) B_1 + 2 m_{n_i}^2 m_{l_k}^2 B_0 \right\}\,,
\end{eqnarray*}
where $B_{0, 1} = B_{0, 1}(m_{l_m}^2, m_{n_i}^2, m_W^2)$.
\[
\tilde{C}_0(p_2^2, p_1^2, m_1^2, m_2^2, m_3^2) \equiv B_0(p_1^2, m_2^2, m_3^2) + m_1^2 C_0(p_2^2, p_1^2, m_1^2, m_2^2, m_3^2)~.
\]
Notice that we have corrected the global sign of $F_L^{(1)}$, which was a typo in \cite{Arganda:2004bz}.

In all the previous formulas, summation over neutrino indices are understood. These run as
 $i,j=1, ..., 9$ for neutrinos, and $k,m=1,..., 3$, for charged leptons. 
The loop function conventions are as in \cite{loopfunctions1,loopfunctions2,loopfunctions3}.

\bibliographystyle{unsrt}

\end{document}